\documentclass[usenatbib]{mn2e}
\usepackage{amsmath, amssymb, graphics, setspace, bm, graphicx}
\allowdisplaybreaks

\newcommand{\mathsym}[1]{{}}
\newcommand{\unicode}[1]{{}}

\newcommand{\Sg}{\Sigma}
\newcommand{\ag}{\alpha}
\newcommand{\bg}{\beta}
\newcommand{\cg}{\gamma}

\newcommand{\dg}{\delta}
\newcommand{\Dg}{\Delta}
\newcommand{\kg}{\kappa}

\newcommand{\Om}{\Omega}
\newcommand{\om}{\omega}
\newcommand{\pom}{\varpi}
\newcommand{\lam}{\lambda}
\newcommand{\vphi}{\varphi}
\newcommand{\pd}{\partial}
\newcommand{\im}{{\rm i}}
\newcommand{\der}{\text{d}}
\newcommand{\tW}{\tilde W}
\newcommand{\tWs}{\tilde W_\star}

\newcommand{\cs}{c_{\rm s}}

\newcommand{\tbg}{\tilde \beta}
\newcommand{\tphi}{\tilde \varphi}

\newcommand{\bfv}{{\bm f}_{\rm v}}

\newcommand{\kr}{k_r}
\newcommand{\kz}{k_z}

\newcommand{\rin}{r_{\rm in}}
\newcommand{\rout}{r_{\rm out}}

\newcommand{\bl}{{\bm {\hat l}}}

\newcommand{\ls}{{\bm {\hat l}}_\star}
\newcommand{\bs}{{\bm {\hat s}}}

\newcommand{\eps}{\epsilon}
\newcommand{\omh}{\omega_\bullet}
\newcommand{\bomh}{\bar \omega_\bullet}
\newcommand{\bomhr}{\bar \omega_{\bullet, {\rm rigid}}}
\newcommand{\bomkg}{\bar \omega_\kappa}
\newcommand{\bcgv}{\bar \gamma_{\rm v}}

\newcommand{\rkg}{r_\kappa}
\newcommand{\cgf}{\gamma_{\rm f}}
\newcommand{\bcgf}{\bar \gamma_{\rm f}}

\newcommand{\tkg}{\tilde \kappa}
\newcommand{\tOmp}{\tilde \Omega_\perp}

\newcommand{\tbw}{t_{\rm bw}}

\newcommand{\Ws}{W_\star}
\newcommand{\tG}{\tilde G}
\newcommand{\tGs}{\tilde G_\star}

\newcommand{\Qv}{Q_{\rm v}}

\newcommand{\Qp}{Q_{\rm p}}

\newcommand{\bv}{{\bm v}}

\newcommand{\rg}{\rho}

\newcommand{\ah}{a_\bullet}

\newcommand{\bx}{{\bm {\hat x}}}
\newcommand{\by}{{\bm {\hat y}}}
\newcommand{\bz}{{\bm {\hat z}}}

\newcommand{\dotm}{{\dot m}}

\newcommand{\dotMedd}{{\dot M}_{\rm Edd}}

\newcommand{\Ms}{M_\star}
\newcommand{\Rs}{R_\star}
\newcommand{\Rst}{R_\star}
\newcommand{\Mh}{M_\bullet}
\newcommand{\Rt}{R_{\rm t}}
\newcommand{\dotMfb}{\dot M_{\rm fb}}
\newcommand{\tf}{t_{\rm f}}

\newcommand{\Rg}{R_{\rm g}}

\newcommand{\cH}{\mathcal{H}}

\newcommand{\Msun}{{\rm M}_\odot}
\newcommand{\Rsun}{{\rm R}_\odot}

\newcommand{\rc}{r_{\rm c}}

\newcommand{\btimes}{{\bm \times}}
\newcommand{\bcdot}{{\bm \cdot}}

\newcommand{\rp}{r_{\rm p}}

\newcommand{\cO}{\mathcal{O}}

\newcommand{\be}{\begin{equation}}
\newcommand{\ee}{\end{equation}}

\begin{document}

\title[TDE Disk Warp and Inclination Evolution]
{Tidal Disruption Event Disks around Supermassive Black Holes: Disk Warp and Inclination Evolution}

\author[J. J. Zanazzi and Dong Lai]{J. J. Zanazzi$^{1,2}$\thanks{Email: jzanazzi@cita.utoronto.ca}, and Dong Lai$^{2}$ \\
$^{1}$Canadian Institute for Theoretical Astrophysics, University
of Toronto, 60 St. George Street, Toronto, Ontario, M5S 1A7,
Canada \\
$^{2}$Cornell Center for Astrophysics, Planetary Science, Department of Astronomy, Cornell University, Ithaca, NY 14853, USA}



\maketitle
\begin{abstract}
After the Tidal Disruption Event (TDE) of a star around a SuperMassive Black Hole (SMBH), the bound stellar debris rapidly forms an accretion disk.  If the accretion disk is not aligned with the spinning SMBH's equatorial plane, the disk will be driven into Lense-Thirring precession around the SMBH's spin axis, possibly affecting the TDE's light curve.  We carry out an eigenmode analysis of such a disk to understand how the disk's warp structure, precession, and inclination evolution are influenced by the disk's and SMBH's properties.  We find an oscillatory warp may develop as a result of strong non-Keplarian motion near the SMBH.  The global disk precession frequency matches the Lense-Thirring precession frequency of a rigid disk around a spinning black hole within a factor of a few when the disk's accretion rate is high, but deviates significantly at low accretion rates.  Viscosity aligns the disk with the SMBH's equatorial plane over timescales of days to years, depending on the disk's accretion rate, viscosity, and SMBH's mass.  We also examine the effect of fall-back material on the warp evolution of TDE disks, and find that the fall-back torque aligns the TDE disk with the SMBH's equatorial plane in a few to tens of days for the parameter space investigated.  Our results place constraints on models of TDE emission which rely on the changing disk orientation with respect to the line of sight to explain observations.
\end{abstract}



\section{Introduction}
\label{sec:Intro}

When a star wanders too close to a SuperMassive Black Hole (SMBH) at the center of a galaxy, the tidal force exerted on the star by the SMBH overcomes the star's self-gravity, and the star tidally disrupts. Such tidal disruption events (TDEs) are expected to produce distinct electromagnetic flares \citep{Rees(1988)}: half of the stellar debris escapes from the SMBH on an unbound orbit, while the other half remains gravitationally bound to the SMBH.  This bound material rains down onto the SMBH at a characteristic accretion rate $\dot M \propto t^{-5/3}$, and forms an accretion disk after eccentric fluid streams collide with one another \citep{Rees(1988),EvansKochanek(1989),Cannizzo(1990)}.  This TDE disk proceeds to accrete rapidly onto the SMBH, producing a luminous flare over a few months to years proportional to the fall-back material accreted onto the disk.
 
Over the last few decades, dozens of TDEs or TDE candidates have been discovered in various spectral bands, ranging from soft X-rays (e.g. \citealt{Bade(1996),KomossaBade(1999),Greiner(2000),Maksym(2010),Donato(2014),Maksym(2014),KhabibullinSazonov(2014),Lin(2015)}), hard X-rays \citep{Bloom(2011),Burrows(2011),Levan(2011),Zauderer(2011),Cenko(2012b),Brown(2015)}, to UV (e.g. \citealt{Gezari(2006),Gezari(2008),Gezari(2009)}) and optical (e.g. \citealt{Komossa(2008),vanVelzen(2011),Wang(2011),Wang(2012),Arcavi(2014),Chornock(2014)}).  Ongoing and future transient surveys like ASAS-SN, PTF, Pan-STARRS, ZTF, and LSST are poised to discover and characterize many more TDEs in the coming decade.
 
 Various models attribute TDE emission arising from inefficient circularization of tidal disruption debris \citep{Guillochon(2014),Piran(2015),Shiokawa(2015),Krolik(2016)}, or outflows supported by radiation pressure \citep{LoebUlmer(1997),Bogdanovic(2004),StrubbeQuataert(2009),CurdNarayan(2018)}.  If the outflow absorbs the inner accretion disk's X-ray and ultraviolet emission \citep{MetzgerStone(2016)}, the full range of observed emission from TDEs may be explained by the observer's viewing geometry \citep{Dai(2018)}.  It is often assumed that the TDE disk is parallel to the SMBH's equitorial plane.  But if the star disrupts on an orbit misaligned with the SMBH's spin axis, the disk will be driven into precession from Lense-Thirring torques, and potentially align with the SMBH's spin over longer timescales.

 Some models explaining a TDE's hard X-ray emission invoke a misaligned accretion disk as an essential component.  The spectra of the TDEs Swift J164449.3+573451 \citep{Bloom(2011),Burrows(2011),Levan(2011),Zauderer(2011)}, Swift J2058.4+0516 \citep{Cenko(2012a)}, and Swift J1112.2-8238 \citep{Brown(2015)} were highly non-thermal, implying a jet was contributing to the tidal disruption flare's emission.  Moreover, the light curve of Swift J164449.3+573451 displayed order-of-magnitude quasi-periodic variations in the hard X-ray, with a period of order $\sim 2.7$ days \citep{Burrows(2011),Saxton(2012b)}.  If the TDE disk is misaligned with the spinning SMBH's equitorial plane, the Lense-Thirring effect drives the disk to precess around the SMBH's spin axis \citep{StoneLoeb(2012),Tchekhovskoy(2014),Franchini(2016)}. The jet axis would vary with respect to the observer's line of sight, causing variations in the hard X-ray's light curve.  Some studies assume that the inner edge of the accretion disk is nearly aligned with the SMBH's equatorial plane \citep{Lei(2013)}, while others argue that the entire disk is nearly flat and precesses like a rigid plate around the SMBH \citep{StoneLoeb(2012),ShenMatzner(2014),Franchini(2016)}.  Most studies of misaligned TDE disks assume the only torque aligning the disk with the SMBH's equatorial plane is from the disk's viscosity \citep{StoneLoeb(2012),Lei(2013),Franchini(2016)}.  Recently,  \cite{Ivanov(2018)} included the torque acting on the disk from the stellar debris' fall-back material, and showed that typical TDE disks cannot complete one full precession period before aligning with the SMBH's equatorial plane.

In this work, we attempt to clarify these theoretical issues on the warp profile, precession and inclination dynamics of TDE disks around SMBHs. Setion~\ref{sec:DiskWarp_LT} examines the warp structure and dynamical evolution of a thick ($H/r \gtrsim \ag$) disk with a power-law surface density and constant aspect ratio.  Section~\ref{sec:TDEModel} introduces our simple model TDE disks.  Section~\ref{sec:TDEDisk_LT} contains our results for the precession and damping rates of a viscous TDE disk around a SMBH.  Section~\ref{sec:TDEDisk_FB} investigates how the fall-back material influences the alignment of the TDE disk with the SMBH's equatorial plane.  Section~\ref{sec:Discuss} discusses theoretical uncertainties and observational implications of our work.  Section~\ref{sec:Conc} summarizes our key results.

\section{Warped Disk Undergoing Lense-Thirring Precession}
\label{sec:DiskWarp_LT}

Before considering more detailed models of TDE disks (Sec.~\ref{sec:TDEModel}), in this section we study the warp and dynamical evolution for a simple model of an accretion disk orbiting a Black Hole (BH) of mass $\Mh$ and dimensionless spin parameter $\ah$.  We denote the BH's gravitational radius by $\Rg = G \Mh/c^2$.  We take the disk's surface density $\Sg (r,t) \propto r^{-1/2}$, and the disk aspect ratio $H/r$ to be constant across the disk's annular extent.  The inner truncation radius of the disk is taken to be the Innermost Stable Circular Orbit  (ISCO) of a test particle orbiting prograde around a spinning BH $r_{\rm ISCO}$ \citep{Bardeen(1972)}
\begin{align}
&\rin = r_{\rm ISCO} 
\nonumber \\
&= \left[ 3 + Z_2 - {\rm sgn}(\ah) \sqrt{(3 - Z_1)(3 + Z_1 + 2 Z_2)} \right] \Rg,
\label{eq:rin}
\end{align}
where
\begin{align}
Z_1 &= 1 + (1-\ah^2)^{1/3} \left[ (1+\ah)^{1/3} + (1-\ah)^{1/3} \right], \\
Z_2 &= \sqrt{3\ah^2 + Z_1^2}, \\
{\rm sgn}&(\ah) = 
\left\{ \begin{array}{ll}
+1 & \text{if} \ \ah \ge 0 \\
-1 & \text{if} \ \ah < 0
\end{array} \right.
.
\end{align}
 The outer truncation radius of the disk is set to be $\rout = 94.2 \, \Rg$.

We assume the misalignment between the orbital angular momentum unit vector of the disk $\bl = \bl(r,t)$ and the SMBH's spin vector $\bs$ is small everywhere ($|\bl \btimes \bs| \ll 1$).  Defining the complex warp amplitude $W(r,t) = \bl \bcdot (\bx + \im \by)$ in a Cartesian coordinate system with $\bz = \bs = \bl + \cO(|W|)$ for $\ah > 0$ and $\bz = -\bs = \bl + \cO(|W|)$ for $\ah < 0$, the disk warp evolves in time according to \citep{Ogilvie(1999),LubowOgilvie(2000)}
\begin{equation}
\Sg r^2 \Om \frac{\pd W}{\pd t} = \frac{1}{r} \frac{\pd G}{\pd r} + T,
\label{eq:dWdt}
\end{equation}
where $G(r,t) = G_x(r,t) + \im G_y(r,t)$ is the disk's complex internal torque per unit area, while $T$ is the complex external torque acting on the disk.

Equation~\eqref{eq:dWdt} is closed with an equation for the disk's internal torque, written in terms of the disk warp.  This closure expression depends on whether or not the disk lies in the so-called \textit{resonant} regime \citep{PapaloizouPringle(1983),PapaloizouLin(1995),IvanovIllarionov(1997),Ogilvie(1999),LubowOgilvie(2000)}, which occurs when bending waves are able to travel at approximately half the disk's sound speed.  In the appendix, we derive the dispersion relation for inertial-density waves in viscous disks with non-Keplerian epicyclic frequencies $\kg$.  We show the approximate condition for bending waves to globally propagate across the disk with a velocity $v_{\rm bw} \approx \cs/2$ is
\be
\frac{H}{r} \gtrsim \ag \hspace{5mm} \text{and} \hspace{5mm} \frac{H}{r} \gtrsim \frac{\Om^2 - \kg^2}{2 \Om^2} \equiv \tkg,
\label{eq:res_cond}
\ee
where $H = \cs/\Om$ is the disk scale-height, $\ag$ is the dimensionless Shakura-Sunyaev viscosity parameter, and $\Om$ is the orbital frequency.  When condition~\eqref{eq:res_cond} is satisfied, the disk lies in the resonant regime, with $G(r,t)$ given by \citep{LubowOgilvie(2000)}
\be
\frac{\pd G}{\pd t} = \im \tkg \Om G - \ag \Om G + \frac{\Sg H^2 r^3 \Om^3}{4} \frac{\pd W}{\pd r}.
\label{eq:G_res}
\ee
When condition~\eqref{eq:res_cond} is violated, the disk lies in the \textit{diffusive} regime, and $G(r,t)$ is given by \citep{Ogilvie(1999)}
\be
G = \Sg H^2 r^3 \Om^2 \left( \Qv \frac{\pd W}{\pd r} + \im \Qp \frac{\pd W}{\pd r} \right),
\label{eq:G_visc}
\ee
assuming a steady-state disk (radial velocity $v_r = \frac{3}{2} \ag H^2 \Om/r$).   The viscous and pressure coefficients $\Qv$ and $\Qp$ are given by (\citealt{Ogilvie(1999)}, Appendix A5)
\begin{align}
\Qp &= \frac{2\tkg + \ag^2[ 3 + 2(3+\tkg)(2\tkg-\ag^2)]}{2[(2\tkg-\ag^2)^2 + 4\ag^2]} + \cO(|W|^2), 
\label{eq:Qp} \\
\Qv &= \frac{\ag[1 + 2\tkg + \ag^2(7 + 2\tkg)]}{(2\tkg-\ag^2)^2 + 4 \ag^2} + \cO(|W|^2).
\label{eq:Qv}
\end{align}
Although the dependence of $\Qv$ and $\Qp$ on $\ag$ and $\tkg$ is complicated, two limiting cases are particularly relevant for astrophysical disks.  The first is the high viscosity limit ($|\tkg| \ll \ag^2 \ll 1$), where $\Qv$ and $\Qp$ reduce to (assuming $|W| \ll 1$)
\be
\Qp \simeq \frac{3}{8} \hspace{5mm} \text{and} \hspace{5mm} \Qv \simeq \frac{1}{4\ag}.
\label{eq:Q_highv}
\ee
Clearly for $\ag \lesssim 1$, viscosity is the main internal torque working to maintain the disk's coplanarity ($\Qv \gg \Qp$).  The opposing limit is the low viscosity regime ($\ag^2 \ll |\tkg| \ll 1$), where $\Qp$ and $\Qv$ reduce to (assuming $|W| \ll 1$)
\be
\Qp \simeq \frac{1}{4 \tkg} \hspace{5mm} \text{and} \hspace{5mm} \Qv \simeq \frac{\ag}{4 \tkg^2}.
\label{eq:Q_lowv}
\ee
In the low viscosity limit, pressure can be the main internal torque working to maintain the disk's coplanarity ($\Qp \gg \Qv$).

 The dimensionless function $\tkg$ [Eq.~\eqref{eq:res_cond}] measures the amount of apsidal precession for a slightly eccentric fluid particle's orbit.  Around a Kerr BH, it is given by (e.g. \citealt{Kato(1990)})
\be
\tkg(r) = \frac{3 \Rg}{r} - \frac{4 \ah \Rg^{3/2}}{r^{3/2}} + \frac{3 \ah^2 \Rg^2}{2 r^2}.
\label{eq:tkg}
\ee
Note that $\tkg \sim 1$ in the inner region of the disk.

Most previous studies of disk warps in the diffusive regime around spinning BHs have assumed the disk lies in the high viscosity limit and used Equation~\eqref{eq:Q_highv}, thus essentially neglecting $\tkg$ (see e.g. \citealt{KumarPringle(1985),ScheuerFeiler(1996),LodatoPringle(2006),Martin(2009),TremaineDavis(2014),ChakrabortyBhattacharyya(2017)}).  Such studies find the inner disk to be closely aligned with the spinning BH's equatorial plane (the Bardeen-Peterson effect; \citealt{BardeenPetterson(1975)}).  Hydrodynamical simulations have reproduced this result in the $\ag \gtrsim H/r$ regime, some by neglecting apsidal precession induced by the BH (e.g. \citealt{NelsonPapaloizou(2000),Sorathia(2013),KrolikHawley(2015),Hawleykrolik(2018),Liska(2018b)}). However, ignoring apsidal precession may neglect important features of the disk's warp profile, since warps induce radial pressure gradients in the disk (e.g. \citealt{IvanovIllarionov(1997),LodatoPringle(2007)}), and the epicyclic frequency can become highly non-Keplerian ($\tkg \sim 1$) near the ISCO.  Some analytic works (e.g. \citealt{IvanovIllarionov(1997),Lubow(2002),King(2005),Zhuravlev(2014)}) and hydrodynamical simulations (e.g. \citealt{FragileAnninos(2005),Fragile(2007),Zhuravlev(2014),Morales(2014),Nealon(2016),Liska(2018)}) of disks in the $\ag \lesssim H/r$ limit which include apsidal precession find a very different picture, with the inner disk highly misaligned with respect to the BH's equatorial plane.   As we show below, this behavior arises from the non-negligible influence of pressure torques when the non-Keplerian epicyclic frequency near the BH is properly taken into account.  These effects are relevant for TDE disks in particular since they are likely to be in the $H/r \gtrsim \alpha$ regime.

Since $\tkg$ is non-negligible around a BH, thin ($H/r \ll 1$) disks have some radial extent where the resonant condition~\eqref{eq:res_cond} is violated.  We define the radius $\rkg$ via
\be
\left. \frac{H}{r} \right|_{r = \rkg} = \tkg(\rkg).
\label{eq:rkg}
\ee
Because of the non-monotonic behavior of $\tkg$, in this section we assume $\rkg$ has a single value, taken as a free parameter.  The internal torque $G$ is given by Equation~\eqref{eq:G_res} when $r > \rkg$, and Equation~\eqref{eq:G_visc} when $r \le \rkg$.  We assume the external torque on the disk is the Lense-Thirring torque \citep{Kato(1990)}:
\be
T = \im \Sg r^2 \Om \omh W,
\label{eq:T_LT}
\ee
where
\be
\Om(r) = \frac{c}{\Rg} \left( \frac{r^{3/2}}{\Rg^{3/2}} + \ah \right)^{-1}
\label{eq:Om}
\ee
is the orbital frequency, and
\be
\omh(r) = \Om \left( \frac{2 \ah \Rg^{3/2}}{r^{3/2}} - \frac{3 \ah^2 \Rg^2}{2 r^2} \right)
\label{eq:omh}
\ee
is the Lense-Thirring precession frequency.

To solve Equations~\eqref{eq:dWdt}, \eqref{eq:G_res}, and~\eqref{eq:G_visc}, we look for solutions of the form
\begin{align}
W(r,t) &= \tW(r) e^{\int^t \lam \der t'},
\label{eq:W_decomp} \\
G(r,t) &= \tG(r) e^{\int^t \lam \der t'},
\label{eq:G_decomp}
\end{align}
where in general the complex eigenfrequency $\lam = \lam(t')$ can change in time with the disk properties and external torque.  In this section however, $\lam$ is constant.  Equations~\eqref{eq:W_decomp}-\eqref{eq:G_decomp} assume the background disk properties change over timescales much longer than the precession/damping time $|\lam|^{-1}$, which we will check in later sections.  Relaxing this assumption forces the amplitudes $\tW$ and $\tG$ to depend on $t$ as well as $r$.  We assume a zero-torque boundary condition
\be
\tG(\rin) = \tG(\rout) = 0.
\ee
Because the solutions are linear in $G$ and $W$, we are free to choose a normalization condition for $\tW$ and $\tG$, which we take to be the disk warp at the outer truncation radius $\tW(\rout)$. 

Equation~\eqref{eq:dWdt} can be integrated over $W^* r \der r$ ($W^*$ is the complex conjugate of $W$) to obtain (after integration by parts) an integral expression for the complex eigenfrequency $\lambda = \gamma + \im \omega$, with
\be
\om = \bomh + \bomkg, \hspace{5mm} \cg = \bcgv,
\label{eq:freqs_LT}
\ee
where
\begin{align}
\bomh &= \frac{1}{L_-}\int_{\rin}^{\rout} \Sg r^3 \Om \omh |\tW|^2 \der r,
\label{eq:bomh} \\
\bomkg &= - \frac{1}{L_-} \int_{\rin}^{\rout} \frac{\tilde Q_{\rm p} |\tG|^2}{\Sg H^2 r^3 \Om^2} \der r,
\label{eq:bomkg} \\
\bcgv &=- \frac{1}{L_+} \int_{\rin}^{\rout} \frac{\tilde Q_{\rm v} |\tG|^2}{\Sg H^2 r^3 \Om^2} \der r,
\label{eq:bcgv} \\
L_\pm &= \int_{\rin}^{\rout} \Sg r^3 \Om |\tW|^2 \der r \pm \int_{\rc}^{\rout} \frac{4 |\tG|^2}{\Sg H^2 r^3 \Om^3} \der r,
\label{eq:L_pm}
\end{align}
where $\rc = \max(\rin,\rkg)$, and
\begin{align}
\tilde Q_{\rm p} &=
\left\{ \begin{array}{cl}
4\tkg & r \ge \rkg \\
\Qp/(\Qp^2 + \Qv^2) & r < \rkg
\end{array} \right.  , \\
\tilde Q_{\rm v} &=
\left\{ \begin{array}{cl}
4 \ag & r \ge \rkg \\
Q_{\rm v}/(\Qp^2 + \Qv^2) & r < \rkg
\end{array} \right.  .
\end{align}

Notice that the viscous damping $\bcgv$ always aligns the disk with the BH's equitorial plane ($\bcgv \le 0$), regardless of the the BH's spin direction (sign of $\ah$).  This contrasts the work of \cite{ScheuerFeiler(1996)}, who found that a disk with a small initial misalignment with a retrograde BH has an exponentially growing tilt.  The reason for this discrepancy is that \cite{ScheuerFeiler(1996)} only considered the viscous back-reaction torque on the BH from the disk, a reasonable assumption when the BH spin angular momentum ($|S| = |\ah| \Mh \Rg c$) is much smaller than the disk's orbital angular momentum ($L_{\rm disk} \sim 2\pi \Sg r^2 \Om|_{r=\rout}$).  Since TDE disks lie in the opposite regime ($|S| \gg L_{\rm disk}$), we only need to consider viscous torques acting on the disk.

When the disk is rigidly precessing around the BH ($|\tW| = \text{constant}$, $|\tG| = 0$), the  global precession rate reduces to the rigid-body precession frequency $\om = \bomhr$, where
\be
\bar \omega_{\bullet, {\rm rigid}} = \frac{ \int_{\rin}^{\rout} \Sg r^3 \Om \omh \der r }{ \int_{\rin}^{\rout} \Sg r^3 \Om \der r}.
\label{eq:bomh_rigid}
\ee
Equation~\eqref{eq:bomh_rigid} is often used to estimate the precession rate of a disk around a spinning BH (e.g. \citealt{Fragile(2007),StoneLoeb(2012)}).  In addition, the simple estimate \citep{Bate(2000)}
\be
\cg_{\rm Bate} = -\left. \ag \left( \frac{r}{H} \right)^2 \frac{\bomhr^2}{\Om} \right|_{r=\rout}
\label{eq:cgBate}
\ee
is often used as an approximation to the disk's viscous damping rate (e.g. \citealt{FoucartLai(2014),Franchini(2016)}).  We will investigate the validity of this approximation in later sections.

Although the eigenfrequency is given by the integral expressions above [Eqs.~\eqref{eq:bomh}-\eqref{eq:L_pm}], in practice it must be determined numerically alongside the eigenfunctions $\tW$ and $\tG$.  We use a shooting algorithm \citep{Press(2002)} written in C++ to calculate the warped disk's eigenmodes.  Our code calculates the lowest order eigenmode, since higher order eigenmodes have higher viscous damping rates, and are less important for the tilt's long-term evolution.

\subsection{Disk Warp Profile}
\label{sec:LT_warp}

\begin{figure}
\centering
\includegraphics[scale=0.8]{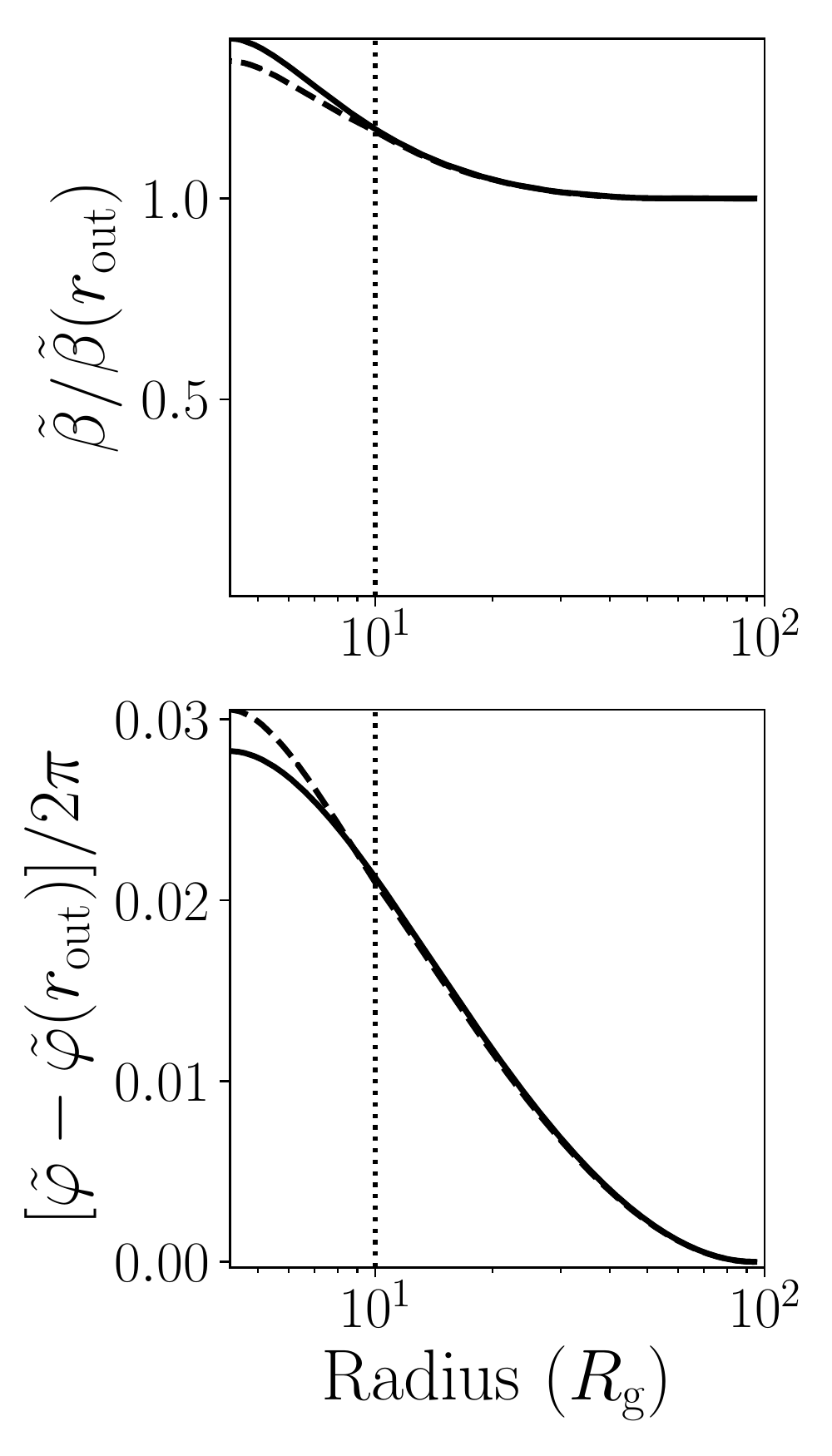}
\caption{Warp $\tilde \bg(r)$ (top) and twist $\tilde \varphi(r)$ (bottom) profiles (defined as $\tW = \tilde \bg e^{\im \tilde \varphi}$) for $\rkg = 0$ (solid) and $\rkg = 10 \Rg$ (dashed).  Here, $\ah = 0.5$, $\ag = 0.1$, and $H/r = 0.5$.  The vertical dotted line marks $\rkg = 10 \, \Rg$.  The Eigenvalues are $\om = 0.056 \, \Om(\rout)$, $\gamma = -0.0067 \, \Om(\rout)$ for $\rkg = 0$, and $\om = 0.055 \, \Om(\rout)$, $\cg = -0.0068 \, \Om(\rout)$ for $\rkg = 10 \, \Rg$ [see Eq.~\eqref{eq:rkg}].}
\label{fig:Warp_LT_ah=0.5}
\end{figure}

\begin{figure}
\centering
\includegraphics[scale=0.8]{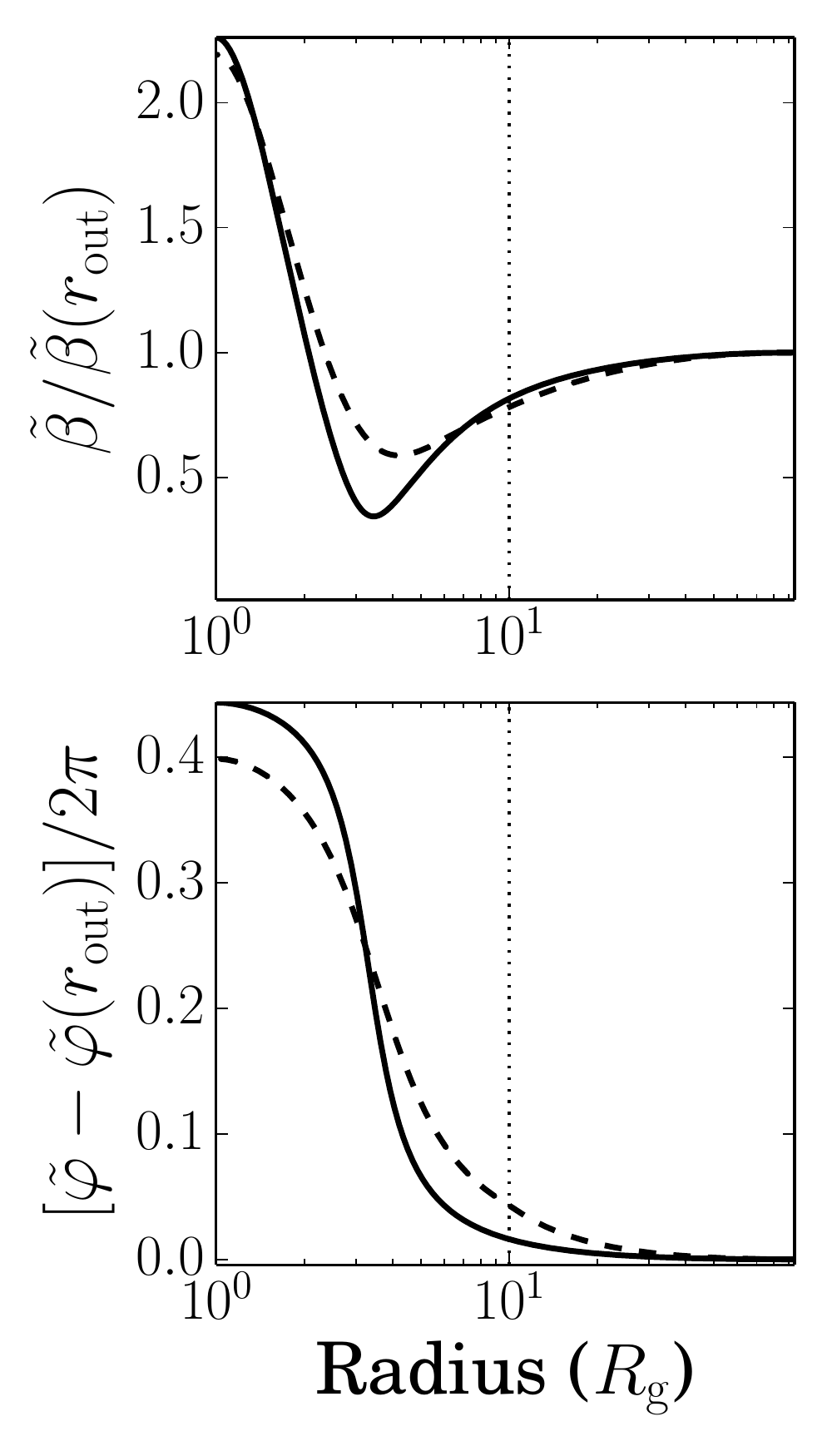}
\caption{ Same as Fig.~\protect\ref{fig:Warp_LT_ah=0.5}, except $\ah = 1.0$.  The Eigenvalues are $\om = -0.0054 \, \Om(\rout)$, $\gamma = -0.044 \, \Om(\rout)$ for $\rkg = 0$, and $\om = 0.0015 \, \Om(\rout)$, $\cg = -0.080 \, \Om(\rout)$ for $\rkg = 10 \, \Rg$.  See Eq.~\eqref{eq:rkg} and discussion thereafter for definition of $\rkg$  }
\label{fig:Warp_LT_ah=1.0}
\end{figure}

\begin{figure}
\centering
\includegraphics[scale=0.8]{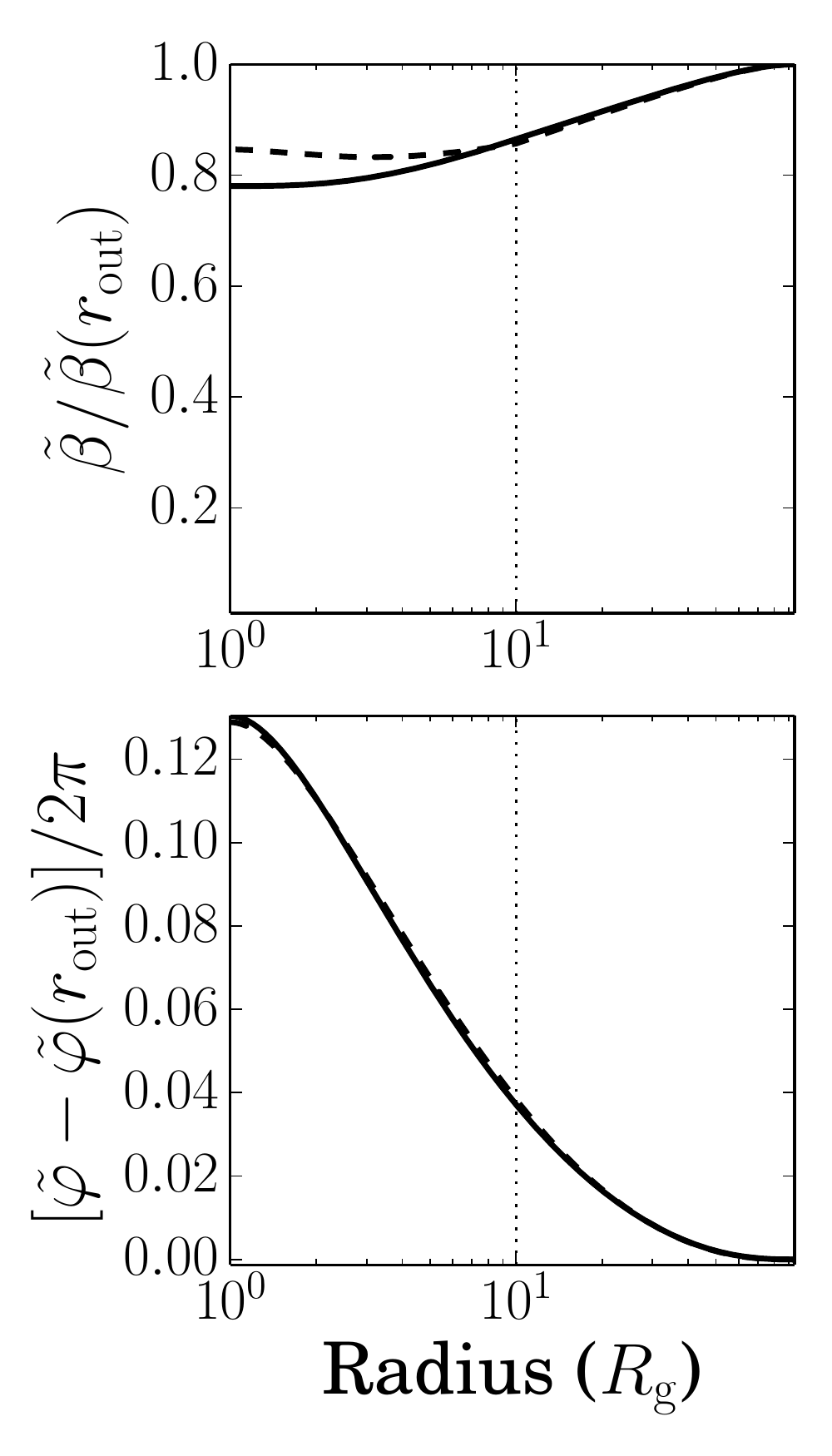}
\caption{Same as Fig.~\protect\ref{fig:Warp_LT_ah=1.0}, except that we set $\tkg = 0$.  The Eigenvalues are $\om = 0.10 \, \Om(\rout)$, $\gamma = -0.045 \, \Om(\rout)$  for $\rkg = 0$, and $\om = 0.10 \, \Om(\rout)$, $\cg = -0.048 \, \Om(\rout)$ for $\rkg = 10 \, \Rg$.}
\label{fig:Warp_LT_nokg}
\end{figure}

\begin{figure}
\centering
\includegraphics[scale=0.8]{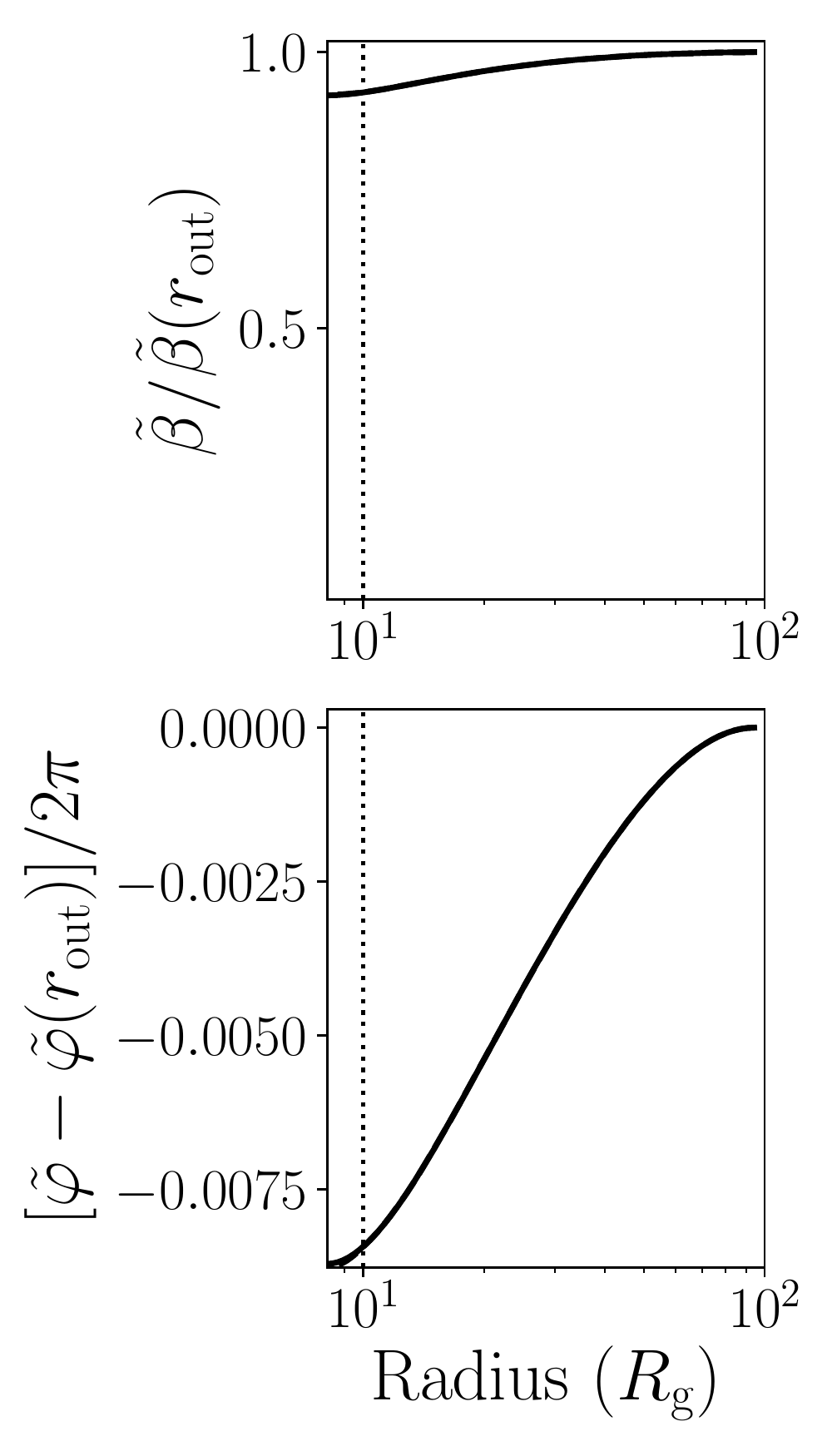}
\caption{ Same as Fig.~\protect\ref{fig:Warp_LT_ah=0.5}, except $\ah = -0.7$.  The Eigenvalues are $\om = -0.031 \, \Om(\rout)$, $\gamma = -0.00095 \, \Om(\rout)$ for $\rkg = 0$ and $\rkg = 10 \, \Rg$.  Notice the $\rkg=0$ (solid) and $\rkg = 10 \, \Rg$ (dashed) solutions lie on top of one another.}
\label{fig:Warp_LT_ah=-0.7}
\end{figure}

Figure~\ref{fig:Warp_LT_ah=0.5} plots the disk warp $\tbg(r)$ (top panel) and twist $\tphi(r)$ (bottom panel) profile for the complex warp amplitude $\tW = \tbg e^{\im \tphi}$ around a BH with a  prograde and moderate spin ($\ah = 0.5$).  The disk warp is close to flat over most of the disk's radial extent, but there is an increase in the warp amplitude $\tbg$ near the disk's inner edge.  This increase is due to the influence of pressure torques, and persists even when the inner disk is in the diffusive regime (dashed line solution with $\rkg = 10 \, \Rg$).  A small twist $\tphi$ develops due to viscous torques.  Notice there are only minor differences between the disk warp profile when the entire disk is in the resonant regime (solid lines), and when the inner disk lies in the diffusive regime (dashed lines), because the dependence of the pressure coefficient $\Qp$ on the dimensionless non-Keplerian epicyclic frequency $\tkg$ has correctly been accounted for.

Figure~\ref{fig:Warp_LT_ah=1.0} is the same as Figure~\ref{fig:Warp_LT_ah=0.5}, except the BH is prograde and spinning maximally ($\ah = 1.0$).  Because the disk is much more extended around such a BH ($\rout \approx 100 \, \rin$ when $\ah = 1$), the disk tilt $\tbg$ oscillations become more pronounced, so that $\tbg(\rin) >2 \, \tbg(\rout)$.  The differences between the $\rkg = 0$ and $\rkg = 10 \, \Rg$ cases are minor, indicating that tilt oscillations occur around maximally spinning BHs when pressure torques are properly taken into account, no matter what regime (resonant or diffusive) the disk lies in (assuming $\ag \lesssim H/r$).  The disk twist $\tphi$ increases rapidly when the disk warp $\tbg$ is small, which can be shown to be due to a coordinate singularity when $\tbg = 0$.  The gradual increase in disk tilt $\tphi$ before and after the dip in disk warp $\tbg$ is due to viscous torques.

In contrast to Figure~\ref{fig:Warp_LT_ah=1.0}, Figure~\ref{fig:Warp_LT_nokg} shows that when $\tkg$ is neglected, the efficacy of pressure torques to drive tilt oscillations is reduced.  We see that the disk warp $\tbg$ decreases smoothly and monotonically to the inner disk edge.  This suggests the Bardeen-Peterson effect [$\tbg(\rin) \ll \tbg(\rout)$] is only possible when pressure torques are negligible ($\Qp \ll \Qv$).

Figure~\ref{fig:Warp_LT_ah=-0.7} looks at the disk warp $\tbg$ and twist $\tphi$ profiles for a retrograde spinning BH ($\ah = -0.7$).  In contrast to prograde BHs, retrograde BHs have decreasing tilts $\tbg$ as the inner radius $\rin$ is approached.  A retrograde BH disk twists in the opposite direction as a prograde BH disk [$\tphi \le \tphi(\rout)$ near $\rout$ when $\ah < 0$, while $\tphi \ge \tphi(\rout)$ near $\rout$ when $\ah > 0$], as seen in other works (e.g. \citealt{ScheuerFeiler(1996), ZhuravlevIvanov(2011)})

\subsection{Precession/Damping Rates}
\label{sec:LT_freqs}

\begin{figure}
\centering
\includegraphics[scale=0.57]{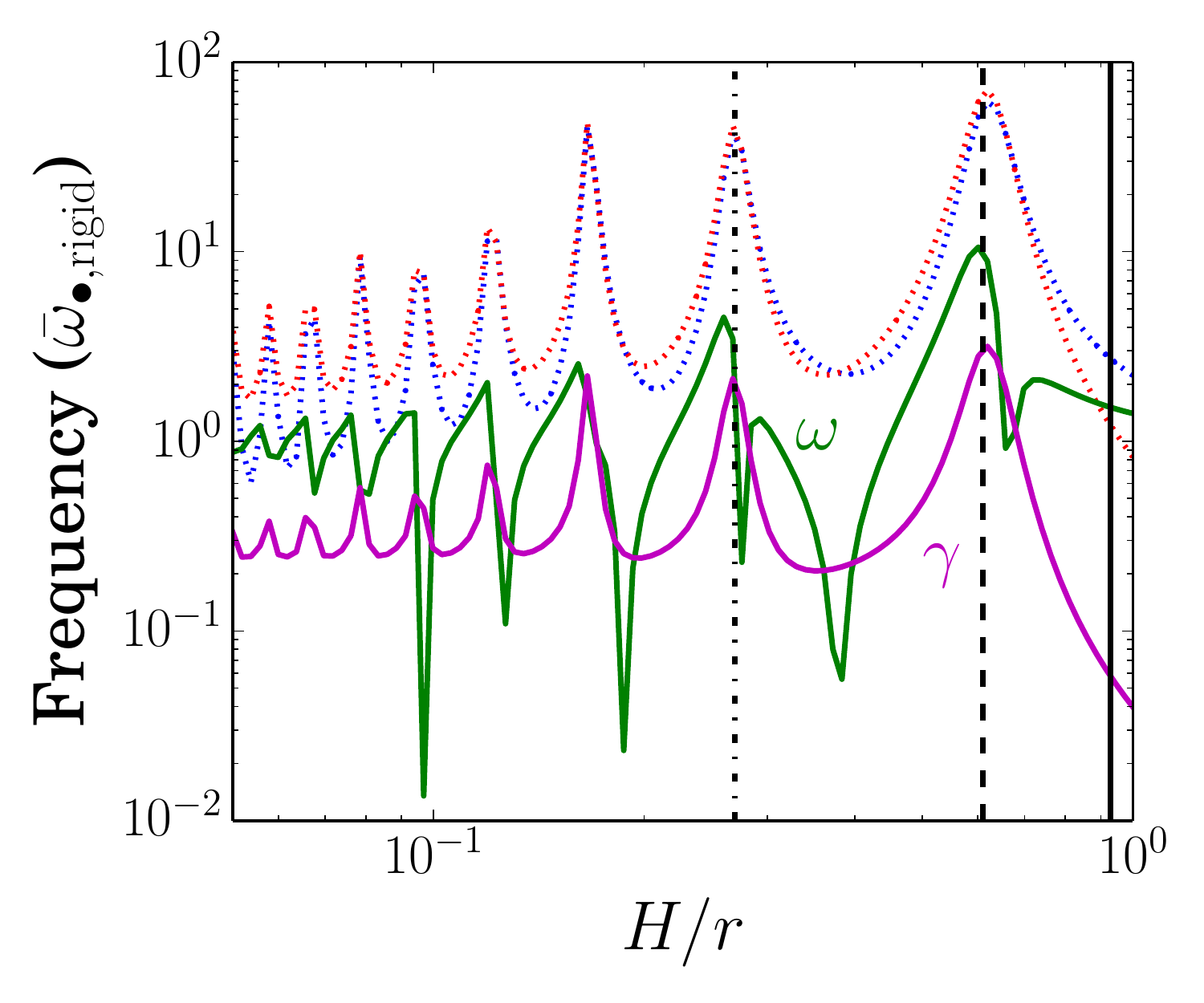}
\caption{Precession frequency $|\om|$ (solid green)  and damping rate $|\cg|$ (solid magenta) as functions of the (constant) disk aspect ratio $H/r$, in units of the rigid body Lense-Thirring precession frequency $|\bar \omega_{\bullet, {\rm rigid}}|$ [Eq.~\eqref{eq:bomh_rigid}].  Dotted lines denote the Lense-Thirring $|\bomh|$ [blue, Eq.~\eqref{eq:bomh}] and non-keplerian epicyclic $|\bomkg|$ [red, Eq.~\eqref{eq:bomkg}] parts of the disk's precession frequency.  Here, $\ah = 1.0$, $\ag = 0.01$, $\rkg = 3 \, \Rg$, and $\Sigma \propto r^{-1/2}$.  Vertical black lines mark $H/r = 0.9$ (solid), $H/r = 0.61$ (dashed), and $H/r = 0.27$ (dot-dashed), with twist/warp profiles plotted in Fig.~\ref{fig:ToyFreq_Sols}.}
\label{fig:ToyFreqs_H/r}
\end{figure}

\begin{figure}
\centering
\includegraphics[scale=0.65]{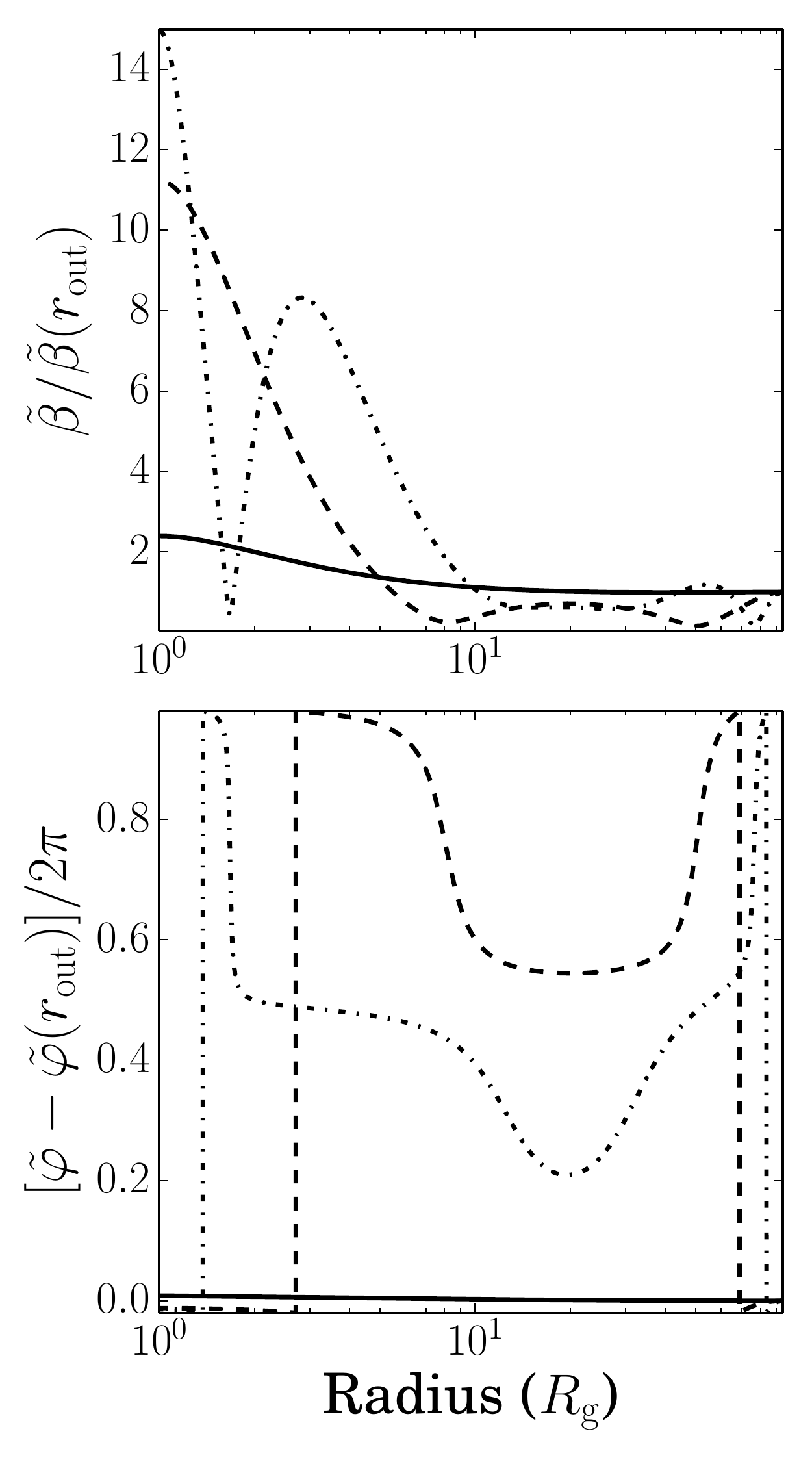}
\caption{Warp $\tbg(r)$ (top panel) and twist $\tphi(r)$ (bottom panel) profiles for $H/r = 0.91$ (solid), $H/r = 0.61$ (dashed), and $H/r = 0.27$ (dot-dashed).  Here, $\ah = 1.0$, $\ag = 0.1$, and $\rkg = 3 \, \Rg$.  See Fig.~\ref{fig:ToyFreqs_H/r} for the precession frequencies and damping rates.}
\label{fig:ToyFreq_Sols}
\end{figure}

Figure~\ref{fig:ToyFreqs_H/r} plots the precession frequency $\om$ and damping rate $\cg$ for a prograde maximally spinning BH ($\ah = 1$), as a function of the disk scale-height $H/r$.  Sharp dips in the precession frequency $\om$ occur when the Lense-Thirring part of the precession frequency $\bomh$ [Eq.~\eqref{eq:bomh}] becomes nearly equal to the non-Keplerian part of the precession frequency $\bomkg$ [Eq.~\eqref{eq:bomkg}].  During these dips, the damping rate $\cg$ becomes larger than the precession frequency $\om$, implying that an initially misaligned disk would rapidly align with the BH's equatorial plane.  Note that the precession frequency $\om$ depends non-trivially on the disk aspect ratio $H/r$, and can differ many orders of magnitude from the naive rigid-body estimate $\bomhr$ [Eq.~\eqref{eq:bomh_rigid}].  
Similar behaviour was seen in \cite{ZhuravlevIvanov(2011)}, where an infinite, steady-state, nearly invicid disk around a spinning BH was considered.  It was found the tilt ratio between the inner and outer edges of the disk $\tbg(\rin)/\tbg|_{r\to \infty}$ diverges for certain discrete values of $H/r$.  We have checked that our tilt eigenmodes $\tbg$ at the locations where $\om$ changes discontinuously have $\tbg(\rin)/\tbg(\rout) \gg 1$.

Figure~\ref{fig:ToyFreq_Sols} plots the warp $\tbg$ (top panel) and twist $\tphi$ (bottom panel) profiles of the complex warp amplitude $\tW = \tbg e^{\im \tphi}$ at several values of $H/r$, denoted by the vertical black lines of Fig.~\ref{fig:ToyFreqs_H/r}.  We see that the spikes in the precession frequency $\om$ coincide with an increase in the number of nodes [where $\tbg(r) \approx 0$] in $\tbg$.  Sharp increases in $\tphi$ occur near warp nodes.  

 Many qualitative features of the $\om$ value's erratic dependence on $H/r$ may be understood by examining the WKB limit of the disk.  As discussed in Appendix~\ref{sec:LT_Toy}, in the WKB approximation the disk warp
\be
\frac{\der ^2 W}{\der r^2} + \frac{V}{r^2} W \simeq 0,
\label{eq:WD_WKB}
\ee
where
\be 
V(r) = \tbw^2 (\om - \omh)(\om - \tkg \Om),
\label{eq:V_eff}
\ee
and $t_{\rm bw} = 2 r/c_{\rm s}$ is the bending wave travel time.  Notice the similarity between Eq.~\eqref{eq:WD_WKB} and the equation for non-radial stellar oscillations in the WKB limit [e.g. \citealt{FullerLai(2012)}, Eq. (16)].  Low-frequency($|\om| < |\omh|, |\tkg \Om|$) disk warp modes would correspond to g-modes of stellar oscillations.  Non-trivial propagating and evanescent zones have been shown to create complex g-mode responses for white dwarf binary stars \citep{FullerLai(2012),FullerLai(2013)}.  Therefore, a complex response is expected of the disk precession frequency to the disk parameters as the disk scale-height (or equivalently $\tbw$) is varied.

Notice the non-trivial propagating and evanescent zones only occur for prograde BH spins ($\ah > 0$).  Because BHs have (global) precession frequencies $\om$ with the same sign as the Lens-Thirring precession frequency $\omh$, retrograde BH spins ($\ah < 0$) always satisfy $\om < \tkg \Om$, so the inner edge of the disk is evanescent to bending waves.  This leads to a decreasing disk tilt as the inner edge of the disk is approached (see Fig.~\ref{fig:Warp_LT_ah=-0.7}), and a less complicated dependence of $\om$ on the disk scale-height.

\section{Tidal Disruption Event Disk Model}
\label{sec:TDEModel}

This section introduces our model for the disk structure formed after a Tidal Disruption Event (TDE) of a star.  We begin by reviewing the physics of TDEs, then follow with our model of TDE disks.  

\subsection{TDE review}
\label{sec:TDEReview}

 A TDE occurs when a star of mass $\Ms$ and radius $\Rs$ approaches a SMBH of mass $\Mh$ on a nearly parabolic orbit, with pericenter distance
\be
\rp \lesssim \Rt = \Rs \left( \frac{\Mh}{\Ms} \right)^{1/3}.
\ee
The energy spread of the star's debris $\Dg E$ after the TDE is (assuming $\Rs \ll \rp$)
\be
\Dg E \simeq \frac{G \Mh \Rs}{\rp^2}.
\ee
Since the star is on a nearly parabolic orbit, the mean energy of the debris $E \approx 0$, so the shortest period of all debris streams is (assuming $\rp \approx \Rt$)
\begin{align}
\tf &= \frac{G\Mh}{(\Dg E)^{3/2}} = \frac{\pi}{\sqrt{2}} \frac{\rp^3}{\sqrt{G \Mh \Rst^3}}
\nonumber \\
& = 41 \left( \frac{\Mh}{10^6 \, \Msun} \right)^{1/2} \left( \frac{1 \, \Msun}{\Ms} \right) \left( \frac{\Rst}{1\,\Rsun} \right)^{3/2} \text{days}.
\label{eq:tf}
\end{align}
 After $\tf$, the streams begin to self-intersect and circularize (\citealt{Rees(1988)}; see Sec.~\ref{sec:ThryUncert} for discussion on circularization efficiency), and an accretion disk forms.  The remaining bound stellar debris then rains down onto the newly formed accretion disk at a rate $\der M_{\rm fb}/\der t$.  Since Kepler's laws give $|E| \propto P^{-2/3}$ ($P$ is the orbital period of the debris stream), the fall-back rate of the bound debris onto the disk is
\be
\frac{\der M_{\rm fb}}{\der t} = \frac{\der M_{\rm fb}}{\der E} \frac{\der E}{\der t} = \frac{\Ms}{3 \tf} \left( \frac{\tf}{t} \right)^{5/3}.
\label{eq:dMfbdt}
\ee
Here, we have assumed the spread in energy of the fall-back material $M_{\rm fb}$ is uniform in $E$ ($\der M_{\rm fb}/\der E \approx \text{constant}$). Analytic arguments and simulations suggest that this approximation breaks down when $t \sim \tf$, but is excellent when $t \gg \tf$ \citep{Lodato(2009),GuillochonRamirez-Ruiz(2013)}.

Normalizing Eq.~\eqref{eq:dMfbdt} to the Eddington accretion rate $\dotMedd = \eta L_{\rm Edd}/c^2$, where $\eta$ is an efficiency factor and $L_{\rm Edd}$ is the Eddington Luminosity, we have
\begin{align}
&\dot m_{\rm fb} = \dotMfb/\dotMedd
\nonumber \\
&= 133.8 \left( \frac{\eta}{0.1} \right) \left( \frac{10^6 \, \Msun}{\Mh} \cdot \frac{1 \, \Rsun}{\Rs} \right)^{3/2}   \left( \frac{\Ms}{1 \, \Msun} \right) \left( \frac{\tf}{t} \right)^{5/3}.
\end{align}
We see almost all TDEs have super-Eddington accretion rates over some portion of their lifetime.

\subsection{Disk Model}
\label{sec:TDEDiskModel}

Our disk model is motivated by \cite{Cannizzo(1990)}, and does not include effects which are important soon after the formation of the TDE disk (e.g. viscous spreading) discussed in \cite{Montesinos(2011)} and \cite{ShenMatzner(2014)}.  We assume the disk forms rapidly after $t \ge \tf$.  The inner truncation radius $\rin$ of the disk is given by the ISCO [Eq.~\eqref{eq:rin}].  The outer truncation radius of the disk $\rout$ is given by the circularization radius of the nearly parabolic debris stream:
\begin{align}
\rout &\approx 2 \Rt = 0.94 \, \text{au} \left( \frac{\Mh}{10^6 \, \Ms} \right)^{1/3} \left( \frac{\Rst}{\Rsun} \right)  \left( \frac{1 \, \Msun}{\Ms} \right)^{1/3} 
\nonumber \\
&= 94.2 \left( \frac{\Rs}{1 \, \Rsun} \right) \left( \frac{10^6 \, \Msun}{\Mh} \right)^{2/3} \left( \frac{1 \, \Msun}{\Ms} \right)^{1/3} \Rg.
\label{eq:rout}
\end{align}

We assume the disk is in a steady state, with an accretion rate $\dot M = -2\pi v_r \Sg \simeq -3\pi \nu \Sg$ that is radially constant ($v_r$ is the radial velocity).  Parameterizing the disk viscosity through the Shakura-Sunyaev $\ag$ prescription ($\nu = \ag H^2 \Om, \ag = \text{constant}$), the viscous heating rate (per unit area) of the disk is
\be
Q_{\rm visc}^+ = \nu \Sg r^2 \left( \frac{\der \Om}{\der r} \right)^2 \simeq \frac{9}{4} \ag \Sg H^2 \Om^3,
\label{eq:Qvisc}
\ee
where the disk scale-height $H$ is related to the isothermal sound-speed $\cs$ via $H = \cs/\Om$.  The disk is cooled by advection and radiation.  The advective cooling rate is \citep{Abramowicz(1988),Abramowicz(1995)}
\be
Q_{\rm adv}^- = \Sg v_r T \frac{\pd s}{\pd r} \approx \frac{9 \Sg \nu \cs^2}{4 r^2},
\label{eq:Qadv}
\ee
where $s$ is the disk entropy, $T$ is the temperature, and we have assumed the constant $\xi$ in \cite{Abramowicz(1995)} is $\xi \approx 3/2$.  The radiative cooling rate of the disk is
\be
Q_{\rm rad}^- = \frac{4 a c T^4}{3 \kg \Sg}.
\ee

We focus on the early phase of the TDE disk, when the disk is supported primarily by radiation pressure ($p \simeq p_{\rm rad} = a T^4/3$).  Since the disk's surface density $\Sg$ is related to the density $\rg$ via $\Sg = 2 H \rg$, the disk sound-speed $\cs$ is given by
\be
\cs^2 = \frac{p}{\rg} \simeq \frac{2 a H T^4}{3 \Sg}
\label{eq:cs_rad}
\ee
Using Equations~\eqref{eq:Qvisc}-\eqref{eq:cs_rad}, energy balance ($Q_{\rm visc}^+ = Q_{\rm adv}^- + Q_{\rm rad}^-$) gives the disk's aspect ratio:
\be
\frac{H}{r} = \sqrt{\cH^2 + 1} - \cH,
\label{eq:H/r}
\ee
where
\be
\cH(r,t) = \frac{4\pi c r}{3 \kg \dot M} = \frac{3.89 \times 10^{-2}}{\dot m} \left( \frac{\eta}{0.1} \right) \left( \frac{r}{\Rg} \right)
\ee
parameterizes the relative importance of advective to radiative cooling in the disk, $\kg = 0.34 \, \text{cm}^2/\text{g}$ is the electron scattering opacity, and
\be
\dot m = \dot M/\dotMedd.
\label{eq:dotm}
\ee
When the disk is advective ($\cH \ll 1$), the disk aspect ratio reduces to
\be
H/r \simeq 1
\label{eq:H/r_adv}
\ee
while when the disk is radiative ($\cH \gg 1$),
\be
H/r \simeq 1/(2 \cH).
\label{eq:H/r_rad}
\ee
In a steady state,
\be
\Sg(r,t) \simeq \frac{\dot M}{3\pi \ag H^2 \Om},
\label{eq:Sigma}
\ee
so $\Sg \propto \dot M r^{-1/2}$ when $\cH \ll 1$, while $\Sg \propto \dot M^{-1} r^{3/2}$ when $\cH \gg 1$.  Notice the disk surface density \textit{increases} with $r$ when the accretion rate becomes sufficiently small.  More detailed models of BH accretion disks give different behaviour of $\Sg$ when the disk is radiative (e.g. \citealt{Ivanov(2018)}).

The scale-height given by Equation~\eqref{eq:H/r} is essentially that derived by \cite{StrubbeQuataert(2009)}, except we do not include the factor $f = 1-\sqrt{\rin/r}$ to force the viscous torque to be zero at $r = \rin$.  The late time behavior of TDEs is better modeled by an accretion disk without the zero-torque boundary condition at the ISCO radius \citep{BalbusMummery(2018)}, and MHD simulations of accretion onto BHs show that the viscous torques do not necessarily vanish at the ISCO radius (e.g. \citealt{Hawley(2000),Hawley(2011)}).

We will have to consider the gas-pressure dominated regime ($p \simeq p_{\rm gas} = \rg k T/\mu m_{\rm p}$) when the disk's accretion rate falls below (e.g. \citealt{ShenMatzner(2014)})
\begin{align}
\dot m_{\rm gas} = \ &\left.\frac{\dot M}{\dotMedd} \right|_{p_{\rm rad}=p_{\rm gas},r=\rout}
\nonumber \\
= \ &1.14 \times 10^{-2} \left( \frac{\eta}{0.1} \right) \left( \frac{0.01}{\ag} \cdot \frac{10^6 \, \Msun}{\Mh} \right)^{1/8}
\nonumber \\
 &\times \left( \frac{\Rs}{1 \, \Rsun} \right)^{21/16} \left( \frac{1 \, \Msun}{\Ms} \right)^{7/16}.
\label{eq:dotm_gas}
\end{align}
When $\dot m \lesssim \dot m_{\rm gas}$, the disk scale-height falls from Eq.~\eqref{eq:H/r_rad} to
\begin{align}
\frac{H}{r} = \ & 5.56 \times 10^{-3} \ \dotm^{1/5} \left( \frac{0.1}{\eta} \right)^{1/5}
\nonumber \\
&\times \left( \frac{10^6 \, \Msun}{\Mh} \cdot \frac{0.01}{\ag} \right)^{1/10} \left( \frac{r}{\Rg} \right)^{1/20}.\
\label{eq:H/r_gas}
\end{align}
In reality, the disk scale-height $H$ does not transition smoothly from Equation~\eqref{eq:H/r_rad} to~\eqref{eq:H/r_gas} for our prescription for the disk's viscosity.  Rather, when $p \simeq p_{\rm rad}$ and $Q_{\rm visc}^+ \simeq Q_{\rm rad}^-$, the disk is susceptible to a thermal instability, and oscillates between the two states given by Equations~\eqref{eq:H/r_adv} and~\eqref{eq:H/r_gas} \citep{LightmanEardly(1974),Abramowicz(1988),ShenMatzner(2014),XiangGruess(2016)}.  We therefore take Equation~\eqref{eq:H/r} to be a conservative upper limit to the disk aspect ratio when the disk is radiative.  Observations support the lack of the \cite{LightmanEardly(1974)} instability occuring in TDE accretion disks, since a disk with scale-height~\eqref{eq:H/r_gas} would have emission in the soft X-ray and ultraviolet much fainter than observed \citep{vanVelzen(2018)}.

Our model for a TDE disk assumes a steady-state accretion rate.  This assumption is valid as long as the viscous time $t_{\rm v} = r^2/\nu$ is everywhere much less than the timescale over which the TDE disk evolves [$\Ms/2 \dot M \sim \tf$, see Eq.~\eqref{eq:tf}].  Since
\begin{align}
t_{\rm v}(\rout) = \ & 5.2 \left( \frac{0.01}{\ag} \right) \left( \frac{\rout}{H(\rout)} \right)^2
\nonumber \\
&\times \left( \frac{\Rs}{1 \, \Rsun} \right)^{3/2} \left( \frac{1 \, \Msun}{\Ms} \right)^{1/2} \, \text{days},
\label{eq:tv}
\end{align}
this is a reasonable assumption early in the disk's lifetime when the disk is hot ($H/r \sim 1$), but will break down when the disk has cooled significantly ($H/r|_{r = \rout} \lesssim 0.3$).

\begin{figure}
\centering
\includegraphics[scale=0.5]{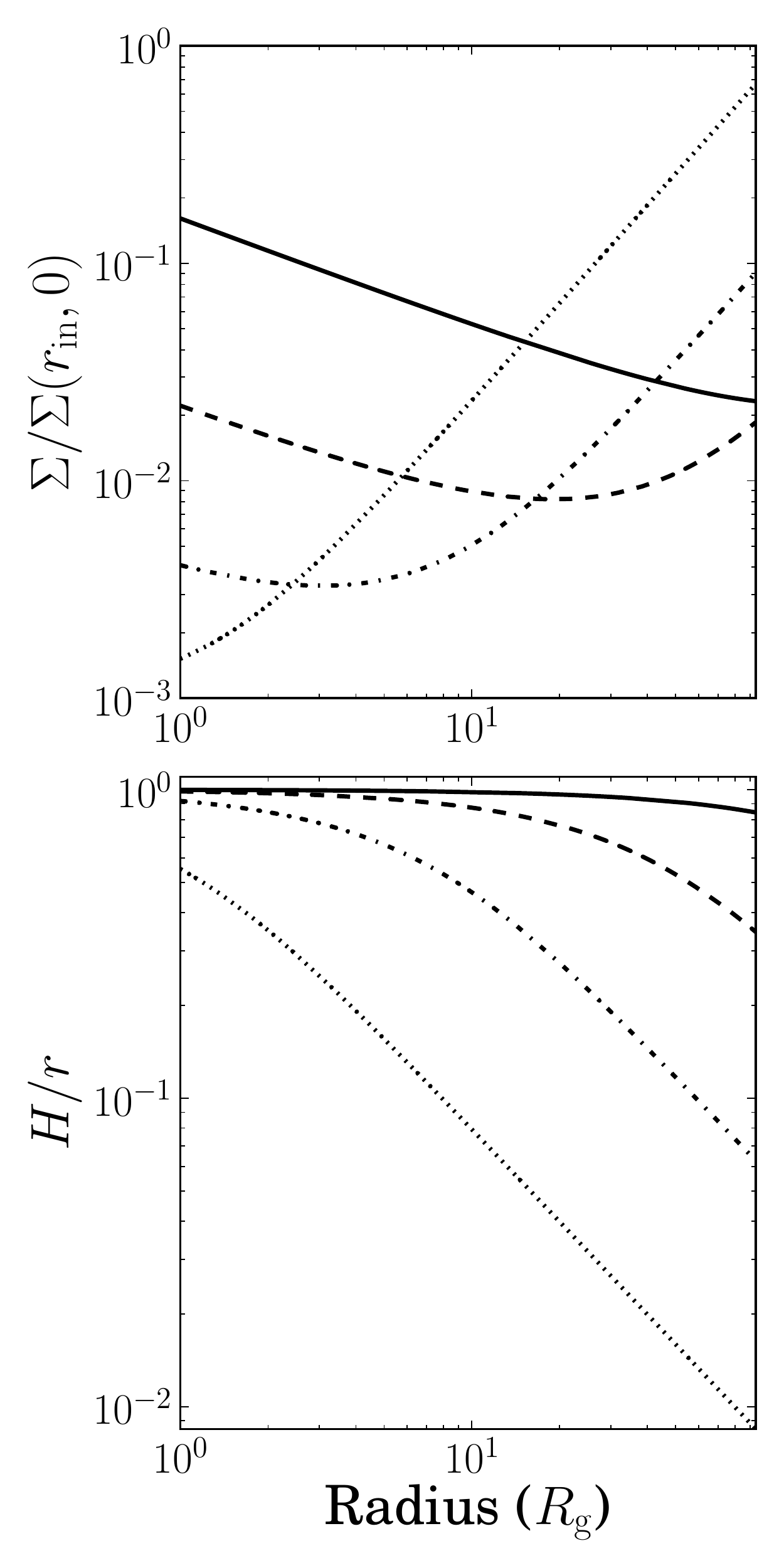}
\caption{Surface density $\Sg$ (top panel) and aspect ratio $H/r$ (bottem panel) of a TDE disk when $\dot m = 21.4$ (solid lines), $\dot m = 2.88$ (dashed lines), $\dot m = 0.462$ (dot-dashed lines), and $\dot m = 0.0624$ (dotted lines).  Here, we take $\Mh = 10^6 \, \Msun$, $\Ms = 1 \, \Msun$, and $\Rs = 1 \, \Rsun$.}
\label{fig:Disk_Profile}
\end{figure}

Figure~\ref{fig:Disk_Profile} shows the surface density profile $\Sg$ and aspect ratio $H/r$ of a TDE disk at different times during its evolution.  At early times, $\Sg \propto r^{-1/2}$ and decreases with the accretion rate $\dot M$.  At later times, the disk begins to cool and the scale-height decreases at a rate proportional to $\dot M$ and becomes $H/r \propto r^{-1}$, while the surface density increases at the disk's outer edges.  The radial profile of the disk's surface density $\Sg$ switches from $\Sg \propto r^{-1/2}$ to $\Sg \propto r^{3/2}$ at these late radiation-cooled stages.

\section{TDE Disk Warp from Lense-Thirring Precession}
\label{sec:TDEDisk_LT}

\begin{figure*}
\centering
\includegraphics[scale=0.5]{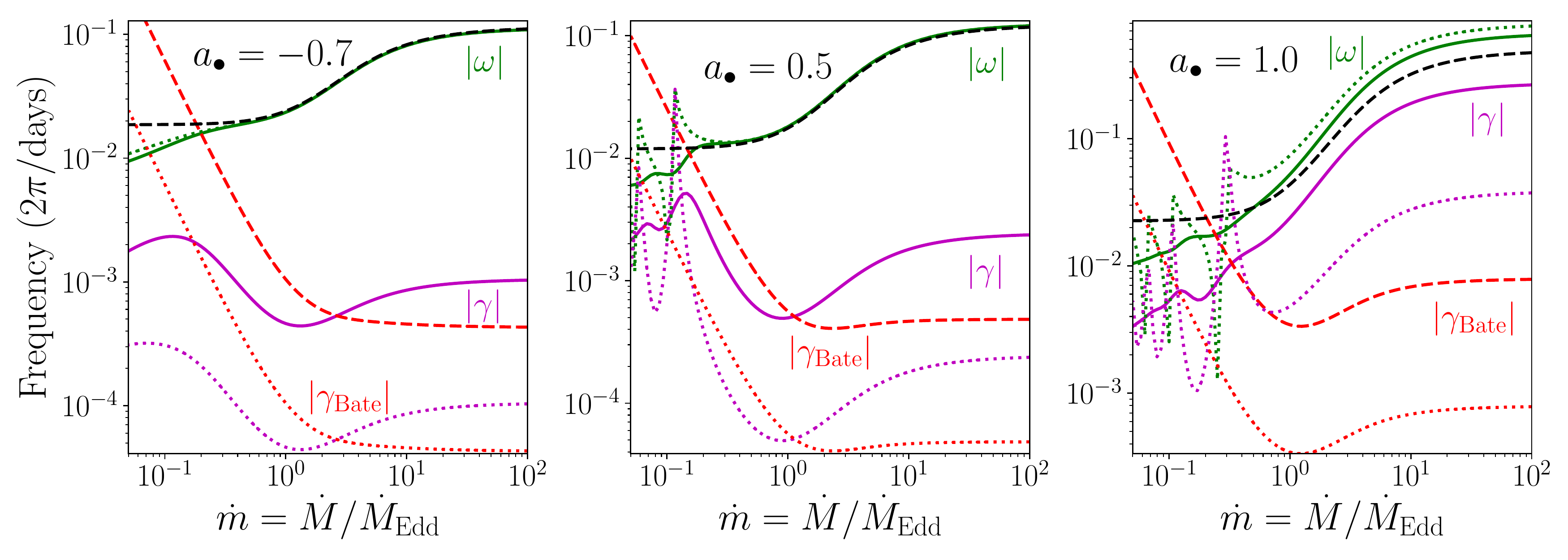}
\caption{ Precession ($\om$, green) and damping ($\cg$, magenta) rates as functions of the disk's accretion rate $\dot m = \dot M/\dotMedd$, for a viscosity parameter of $\ag = 0.1$ (solid) and $\ag = 0.01$ (dotted), with  dimensionless BH spin parameter $\ah$ as indicated.  The black dashed line denotes the rigid-body Lense-Thirring precession frequency $\bar \om_{\bullet, {\rm rigid}}$ [Eq.~\eqref{eq:bomh_rigid}], while red lines denote the \protect\cite{Bate(2000)} viscous damping rate estimate $\cg_{\rm Bate}$ [Eq.~\eqref{eq:cgBate}], for $\ag = 0.1$ (dashed) and $\ag = 0.01$ (dotted). Here, $\Mh = 10^6 \, \Msun$, $\Ms = 1 \, \Msun$, and $\Rs = 1 \, \Rsun$.}
\label{fig:TDE_Freqs}
\end{figure*}

\begin{figure*}
\centering
\includegraphics[scale=0.5]{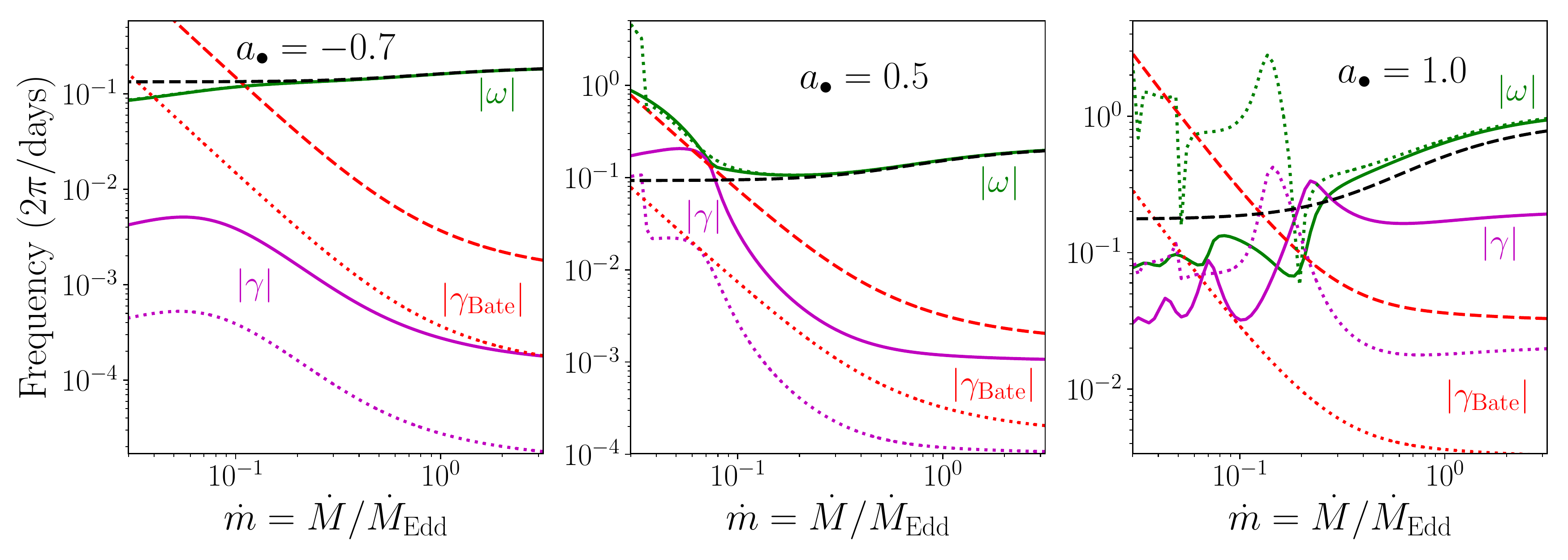}
\caption{ Same as Figure~\protect\ref{fig:TDE_Freqs}, except $\Mh = 10^7 \, \Msun$.}
\label{fig:TDE_Freqs_largeMh}
\end{figure*}

\begin{figure*}
\centering
\includegraphics[scale=0.7]{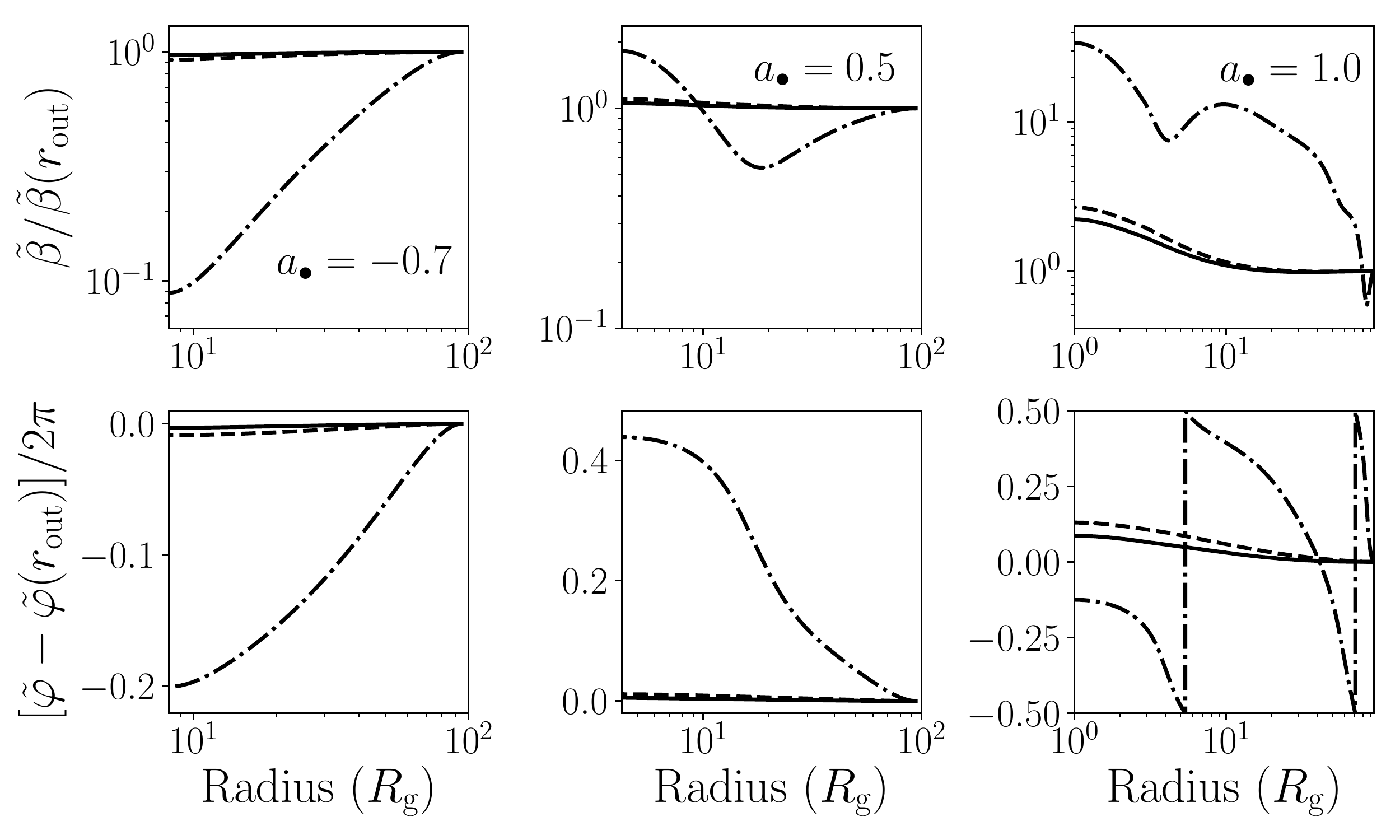}
\caption{ Disk warp $\tbg(r)$ (top panels) and twist $\tphi(r)$ (bottom panels) radial profiles for the complex disk warp eigenfunction $\tW = \tbg e^{\im \tphi}$, for $\dotm = 10$ (solid), $\dotm = 1$ (dashed), and $\dotm = 0.1$ (dot-dashed), with dimensionless SMBH spin parameters $\ah$ as indicated.  Here, $\ag = 0.1$, $\Mh = 10^6 \, \Msun$, $\Ms = 1 \, \Msun$, and $\Rs = 1 \, \Rsun$.  See Figure~\ref{fig:TDE_Freqs} for the precession/damping rates.}
\label{fig:TDE_Warp_highv}
\end{figure*}

\begin{figure*}
\centering
\includegraphics[scale=0.7]{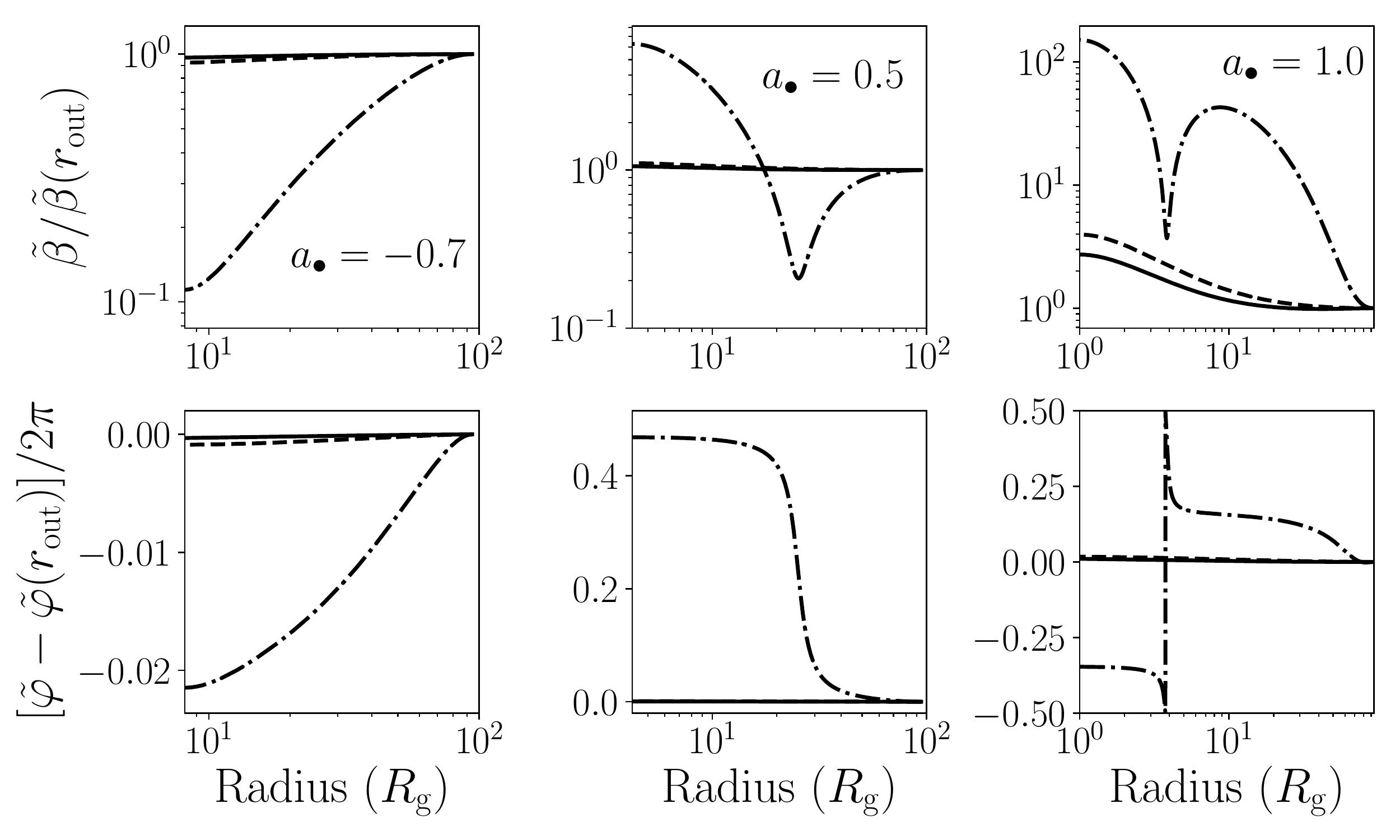}
\caption{ Same as Figure~\ref{fig:TDE_Warp_highv}, except $\ag = 0.01$.  See Figure~\ref{fig:TDE_Freqs} for the precession/damping rates.}
\label{fig:TDE_Warp_lowv}
\end{figure*}

\begin{figure}
\centering
\includegraphics[scale=0.6]{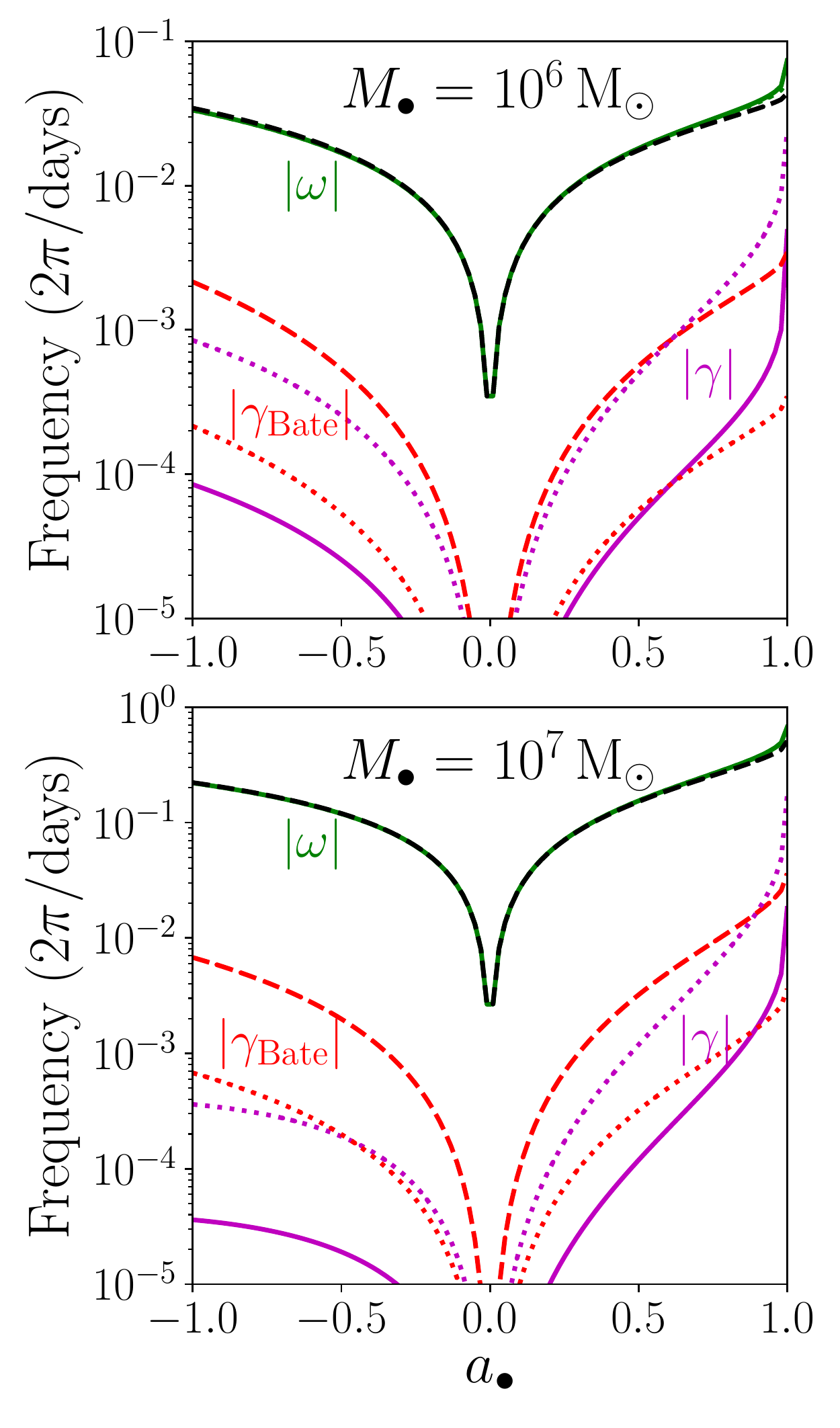}
\caption{ Precession ($\om$, green) and damping ($\cg$, magenta) rates as functions of the SMBH's dimensionless spin parameter $\ah$, for viscosity parameter values of $\ag = 0.1$ (solid) and $\ag = 0.01$ (dotted), and SMBH masses $\Mh$ indicated.  The black dashed line denotes the rigid-body Lense-Thirring precession frequency $\bar \om_{\bullet, {\rm rigid}}$ [Eq.~\eqref{eq:bomh_rigid}], while red lines denote the \protect\cite{Bate(2000)} viscous damping rate estimate $\cg_{\rm Bate}$ [Eq.~\eqref{eq:cgBate}], for $\ag = 0.1$ (dashed) and $\ag = 0.01$ (dotted). Here, $\dot M = \dotMedd$, $\Ms = 1 \, \Msun$, and $\Rs = 1 \, \Rsun$.}
\label{fig:TDE_Freqs_ah}
\end{figure}

This section examines how the TDE disk model described in Section~\ref{sec:TDEDiskModel} evolves its radial warp profile, precession and damping rates in time, due to the Lense-Thirring torque from the SMBH.  As discussed in Section~\ref{sec:LT_freqs}, we expect a non-trivial response of the disk's precession frequency $\om$ to the evolving disk aspect ratio $H/r$.  We solve Equations~\eqref{eq:dWdt}, \eqref{eq:G_res}, and~\eqref{eq:G_visc} numerically using the shooting method written in C++ \citep{Press(2002)} for the eigenfunction $\tW$ and eigenfrequency $\lam = \cg + \im \om$.

Figure~\ref{fig:TDE_Freqs} plots the precession frequency $\om$ and damping rate $\cg$ as a function of the disk's accretion rate $\dotm = \dot M/\dotMedd$.  When the accretion rate is high ($\dotm \gtrsim 0.4$), $\omega$ is always close to (within a factor of 2) the rigid-body Lense-Thirring precession frequency $\bomhr$ estimate [Eq.~\eqref{eq:bomh_rigid}], and $\om$ deviates the most from $\bomhr$ when the BH is spinning prograde rapidly ($\ah \approx 1$).
For low accretion rates ($\dotm \lesssim 0.4$), the dependence of $\om$ on $\dotm$ depends heavily on the disk viscosity $\ag$ and SMBH spin $\ah$.  For high viscosities ($\ag = 0.1$) or retrograde SMBH spins ($\ah = -0.7$), $\om$ decreases below $\bomhr$ as $\dotm$ is lowered.  For low viscosity disks ($\ag = 0.01$) with prograde SMBH spins ($\ah = 0.5, 1.0$), $\om$ suffers significant oscillations as $\dotm$ is decreased, and can differ from $\bomhr$ by more than an order of magnitude.  The viscous damping rates $\cg$ are always at least an order of magnitude below $\om$ for high accretion rates ($\dotm \gtrsim 0.4$), unless the SMBH spin is prograde, near extremal, and viscosity high ($\ah = 1.0, \ag = 0.1$).  Disks around prograde SMBH spins have $\cg$ values comparable to or exceeding $\om$ at low accretion rates ($\dotm \lesssim 0.4$), even when the viscosity is low ($\ag = 0.01$) due to an increase in the number of tilt nodes [when $\tbg(r) \approx 0$] when $\om$ becomes oscillatory (see discussion in Sec.~\ref{sec:LT_freqs} \& App.~\ref{sec:LT_Toy}; see also Figs.~\ref{fig:TDE_Warp_highv}-\ref{fig:TDE_Warp_lowv}), increasing the viscous dissipation in the disk.  Disks orbiting SMBHs with retrograde spins always have $\cg$ values orders of magnitude below their $\om$ values.  The Bate estimate $\cg_{\rm Bate}$ matches the disk's $\cg$ value (within a factor of 10) when the SMBH spin is not prograde and near maximal ($\ah \not\approx 1.0$) and the accretion rate sufficiently high ($\dotm \gtrsim 0.4$), because in these regimes the disk twist becomes non-linear [$|\tphi(r) - \tphi(\rout)| \gtrsim 1$], and $\cg_{\rm Bate}$ implicitly assumes linear disk twists [$|\tphi(r) - \tphi(\rout)| \ll 1$].

Figure~\ref{fig:TDE_Freqs_largeMh} is identical to Figure~\ref{fig:TDE_Freqs}, except we increase the mass of the SMBH to $\Mh = 10^7 \, \Msun$.
The qualitative dependence of the precession $\om$ and damping $\cg$ rates on the disk and SMBH parameters ($\dot m$, $\ag$, and $\ah$) is similar to Figure~\ref{fig:TDE_Freqs}.  The main difference between Figures~\ref{fig:TDE_Freqs} and~\ref{fig:TDE_Freqs_largeMh} is the $\om$ oscillations set in for lower accretion rates ($\dot m \lesssim 0.2$ for Fig.~\ref{fig:TDE_Freqs_largeMh}, $\dotm \lesssim 0.4$ for Fig.~\ref{fig:TDE_Freqs}).

The warp $\tbg(r)$ and twist $\tphi(r)$ radial profiles are plotted for high viscosity eigenfunctions in Figure~\ref{fig:TDE_Warp_highv}, and for low viscosity eigenfunctions in Figure~\ref{fig:TDE_Warp_lowv}, for select accretion rates $\dotm$.  TDE disks orbiting prograde SMBHs typically have inner disks tilted and twisted at larger angles than their outer disks [$\tbg(\rin) > \tbg(\rout)$, $\tphi(\rin) > \tphi(\rout)$ when $\ah > 0$], while TDE disks orbiting retrograde SMBHs generally have inner disks tilted and twisted at smaller angles than their outer disks [$\tbg(\rin) < \tbg(\rout)$, $\tphi(\rin) < \tphi(\rout)$ when $\ah < 0$].  These tilt and twist differences at different disk radii increase as $\dotm$ decreases (with decreasing $H/r$), since the internal torque $G$ maintaining the disk's rigidity is proportional to the disk's scaleheight ($G \propto H$), and becomes less effective when $\dotm$ is small.  Disk twists are higher for larger $\ag$ values because viscous torques become less effective as $\ag$ increases ($G_{\rm visc} \propto \ag^{-1}$, see e.g. \citealt{Ogilvie(1999),Martin(2019)}).  When $\dotm$ becomes sufficiently low ($\dotm \lesssim 0.4$), TDE disks orbiting prograde SMBH spins develop tilt oscillations: $\tbg$ generally decreases in magnitude, but oscillates near inner truncation radius $\rin$.  The $\tphi$ profile changes rapidly near warp nodes.

Figure~\ref{fig:TDE_Freqs_ah} shows how the precession and damping rates of the disk depend on the SMBH spin $\ah$.  At the relatively high accretion rate ($\dot M = \dotMedd$), the rigid-body Lens-Thirring precession frequency $\bomhr$ [Eq.~\eqref{eq:bomh_rigid}] is an excellent approximation to the disk's precession frequency $\om$, and deviates at most by a factor of $\sim 2-3$ only for prograde near-maximally spinning BHs ($\ah \gtrsim 0.9$) due to effects from disk warping.  The viscous damping rate $\cg$ is at least an order of magnitude below $\om$ for the entire rage of viscous parameters $\ag$ considered, unless the SMBH spin is prograde and near maximal ($\ah \gtrsim 0.9$).  The Bate damping rate $\cg_{\rm Bate}$ [Eq.~\eqref{eq:cgBate}] is comparable to $\cg$ (within a factor of 10) unless the SMBH is prograde and spinning rapidly ($\ah \gtrsim 0.9$)

The results of this section use ansatz~\eqref{eq:W_decomp}-\eqref{eq:G_decomp}, which assumes the background disk quantities evolve over a timescale much longer than the precession/damping time $|\lam|^{-1} \sim |\om|^{-1} + |\cg|^{-1}$.  The background quantities evolve fastest when the disk is radiative (when $\dot M \lesssim \dot M_{\rm Edd}$) over the timescale $\Sg/\dot \Sg \sim H/\dot H \sim \tf$ [Eq.~\eqref{eq:tf}; see Sec.~\ref{sec:TDEDiskModel}].  Since $\tf \gtrsim 40 \, \text{days}$ and $|\lam|^{-1} \lesssim 40 \, \text{days}$ for the parameters of interest (see Figs.~\ref{fig:TDE_Freqs}, \ref{fig:TDE_Freqs_largeMh}, \&~\ref{fig:TDE_Freqs_ah}), including the effects of a time-dependent background will not qualitatively affect our results.

\section{Effect of Fall-Back Material}
\label{sec:TDEDisk_FB}

After the star tidally discrupts around the SMBH, the stellar debris rains down on the TDE accretion disk at a rate given by Equation~\eqref{eq:dMfbdt}.  The mass from the fall-back material also deposits angular momentum to the disk, exerting a torque.  This section paramterizes the fall-back torque and examines how the combined influence of Lense-Thirring and fall-back torques affect the disk structure, precession and inclination evolution, using the TDE disk model of Section~\ref{sec:TDEModel}.

Consider a star which disrupts on a parabolic orbit with the orbital angular momentum axis $\ls$.  We parameterize the torque per unit area acting on the TDE disk by
\be
{\bm T}_{\rm f} = \Sg r^2 \Om \cgf \ls,
\label{eq:vTf}
\ee
where
\be
\cgf = \frac{\dotMfb}{2\pi \Sg \rout} \dg(r-\rout)
\label{eq:cgf}
\ee
is the rate of fall-back material accreting onto the outer disk, and $\dg(x)$ is the delta function.  We assume $\ls$ is fixed in time.  For simplicity, we consider the case where $\ls$ does not deviate much from the SMBH spin $\bs$ (the z-axis), and define $\Ws \equiv \ls \bcdot (\bx + \im \by)$.  Thus the complex fall-back torque (per unit area) may be written as
\be
T_{\rm f} = \Sg r^2 \Om \cgf (\Ws - W),
\label{eq:Tf}
\ee
and the total torque (per unit area) acting on the disk is then
\be
T = \im \Sg r^2 \Om \omh W + T_{\rm f}.
\label{eq:T_tot}
\ee

Because of the delta function in Equation~\eqref{eq:cgf}, the fall-back torque $T_{\rm f}$ must be handled with care when included in the warp equations~\eqref{eq:dWdt}, \eqref{eq:G_res}, and~\eqref{eq:G_visc}.  Integrating equation~\eqref{eq:dWdt} over $r \der r$ using the total torque~\eqref{eq:T_tot} from $r = \rout - \eps$ to $r = \rout + \eps$, we see $T_{\rm f}$ causes a discontinuity in the internal torque of
\begin{align}
[G]_{r = \rout} &= \lim_{\eps \to 0} \left( G\big|_{r=\rout + \eps} - G\big|_{r=\rout - \eps} \right) 
\nonumber \\
&= \left. \frac{\dotMfb r^2 \Om}{2 \pi} (\Ws - W) \right|_{r=\rout}.
\label{eq:G_disc}
\end{align}
Requiring $G|_{r = \rout + \eps} = 0$, we see equation~\eqref{eq:dWdt} can be solved with the total torque~\eqref{eq:T_tot} by taking
\be
T = \im \Sg r^2 \Om \omh W
\ee
when $r < \rout$, and forcing $G$ to satisfy the boundary conditions
\be
G\big|_{r=\rin} = 0,
\hspace{5mm}
G\big|_{r=\rout} = \left. \frac{\dotMfb r^2 \Om}{2\pi} (\Ws - W) \right|_{r=\rout}.
\label{eq:G_bdry}
\ee

To solve equations~\eqref{eq:dWdt}, \eqref{eq:G_res}, and~\eqref{eq:G_visc}, we look for solutions of the form
\begin{align}
W(r,t) &= \tW(r) e^{\int^t \lam \der t'} + \tWs(r),
\label{eq:W_decomp_FB} \\
G(r,t) &= \tG(r) e^{\int^t \lam \der t'} + \tGs(r).
\label{eq:G_decomp_FB}
\end{align}
Inserting equations~\eqref{eq:W_decomp_FB}-\eqref{eq:G_decomp_FB} into equations~\eqref{eq:dWdt}, \eqref{eq:G_res}, and~\eqref{eq:G_visc}, we see the functions $\tW$ and $\tG$ satisfy the homogeneous equations
\begin{align}
\frac{\der \tG}{\der r} &= \Sg r^3 \Om (\lam + \im \omh) \tW, & {}
\label{eq:dtGdr} \\
\frac{\der \tW}{\der r} &= \frac{(\lam - \im \tkg \Om + \ag \Om) \tG}{\Sg H^2 r^3 \Om^3} & \text{when } r \ge \rkg,
\label{eq:dtWdr_res} \\
\frac{\der \tW}{\der r} &= \frac{(\Qv - \im \Qp) \tG}{(\Qv^2 + \Qp^2) \Sg H^2 r^3 \Om^2} & \text{when } r < \rkg,
\label{eq:dtWdr_visc}
\end{align}
with the boundary conditions (assuming $\dotMfb = \dot M$)
\be
\tG(\rin) = 0, \hspace{3mm} \tG(\rout) = - \left. \frac{3}{2} \ag \Sg H^2 r^2 \Om^2 \tW \right|_{r=\rout},
\label{eq:tG_bdry}
\ee
while the functions $\tWs$ and $\tGs$ satisfy the equations
\begin{align}
\frac{\der \tGs}{\der r} &= -\im \Sg r^3 \Om \omh \tWs, & {}
\label{eq:dtGsdr} \\
\frac{\der \tWs}{\der r} &= \frac{4(\ag - \im \tkg) \tGs}{\Sg H^2 r^3 \Om^2} & \text{when } r \ge \rkg
\label{eq:dtWsdr_res} \\
\frac{\der \tWs}{\der r} &= \frac{(\Qv + \im \Qp) \tGs}{(\Qv^2 + \Qp^2) \Sg H^2 r^3 \Om^2} & \text{when } r < \rkg,
\label{eq:dtWsdr_visc}
\end{align}
with the boundary conditions (assuming $\dotMfb = \dot M$)
\be
\tGs(\rin) = 0,
\hspace{3mm}
\tGs(\rout) = \left. \frac{3}{2} \ag \Sg H^2 r^2 \Om^2 (\Ws - \tWs) \right|_{r=\rout}.
\label{eq:tGs_bdry}
\ee
Since $\omh \to 0$ as $r \to \infty$, the boundary condition~\eqref{eq:tGs_bdry} is equivalent to $\tWs(\rout) \simeq \Ws$ when $\rout \gg \rin$.  The solutions $\tW$ and $\tG$ evolve in time, precessing and aligning to the SMBH's equatorial plane as the disk evolves.  The solutions $\tWs$ and $\tGs$ do not evolve in time, and correspond to the disk' steady-state profile.  The simulations of \cite{XiangGruess(2016)} looked solely at the steady-state warp profile $\tWs$, while \cite{Ivanov(2018)} simulated the steady-state profile $\tWs$ and the precessing/damping solution $\tW$ simultaneously.

Decomposing the complex eigenfrequency into its real and imaginary parts $\lam = \cg + \im \om$, integrating equation~\eqref{eq:dWdt} over $W^* r \der r$ gives (after integration by parts)
\be
\om = \bomh + \bomkg, 
\hspace{5mm}
\cg = \bcgv + \bcgf,
\label{eq:freqs_FB}
\ee
where $\bomh$ is given by equation~\eqref{eq:bomh}, $\bomkg$ by equation~\eqref{eq:bomkg}, $\bcgv$ by equation~\eqref{eq:bcgv}, and
\be
\bcgf = - \left. \frac{3}{2L_+} \ag H^2 r^2 \Om^2 |\tW|^2 \right|_{r=\rout}
\label{eq:bcgf}
\ee
The modified disk angular momentum $L_+$ is given in equation~\eqref{eq:L_pm}.

Including the fall-back torque $T_{\rm f}$ causes the dynamical part of the disk warp eigenfunction $\tW$ to damp at a faster rate.  After the precessing solution damps to zero ($\tW e^{\int^t \lam \der t'} \to 0$), the disk tilt relaxes to it's steady-state solution ($W \to \tWs$).  Section~\ref{sec:FB_prec} calculates the precession and damping rates for the dynamical disk warp $\tW$.  Section~\ref{sec:FB_steady} calculates how the steady-state $\tWs$ evolves with a changing accretion rate $\dot M$.

\subsection{Dynamical Warp Profiles and Precession/Damping Rates}
\label{sec:FB_prec}

\begin{figure*}
\centering
\includegraphics[scale=0.5]{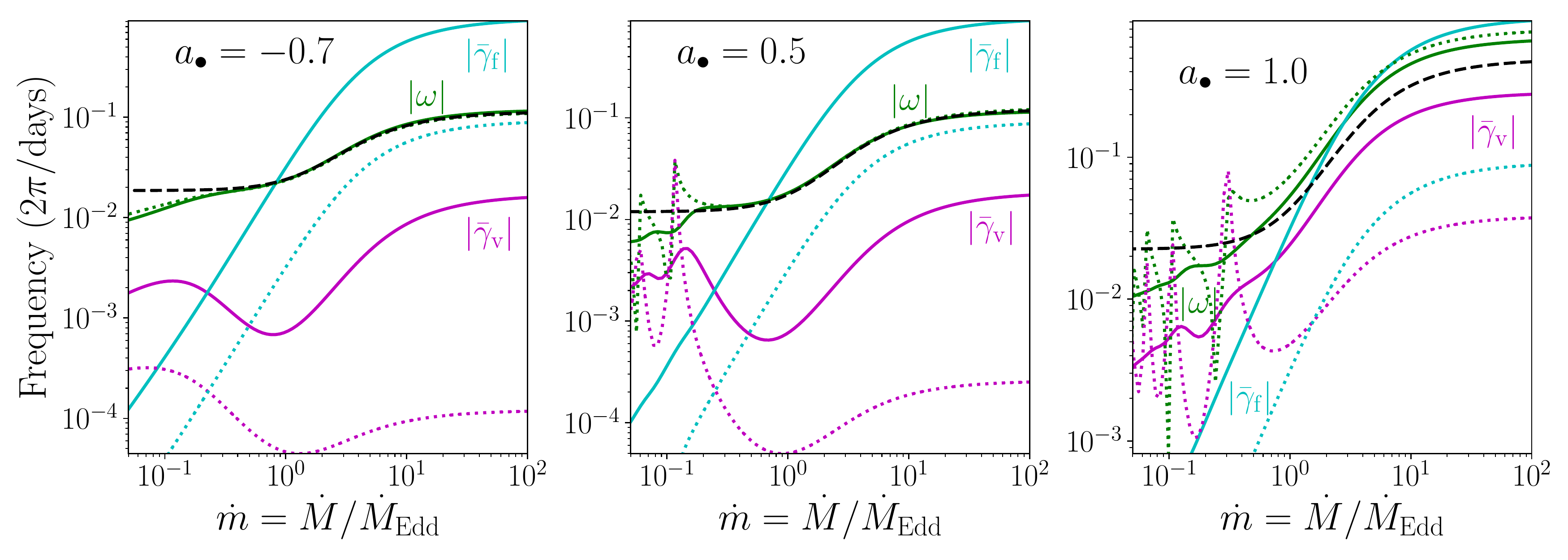}
\caption{ Precession ($\om$, green), viscous damping [$\bcgv$, magenta, Eq.~\eqref{eq:bcgv}], and fall-back damping [$\bcgf$, cyan, Eq.~\eqref{eq:bcgf}] rates as functions of the disk's accretion rate $\dotm = \dot M/\dotMedd$, for viscosity parameters of $\ag = 0.1$ (solid) and $\ag = 0.01$ (dotted), with dimensionless BH spin parameters $\ah$ as indicated.  The black dashed line denotes the rigid-body Lense-Thirring precession frequency $\bomhr$ [Eq.~\eqref{eq:bomh_rigid}] of the disk around the SMBH.  Here, $\Mh = 10^6 \, \Msun$, $\Ms = 1 \, \Msun$, and $\Rs = 1 \, \Rsun$.  We assume $\dot M = \dotMfb$.}
\label{fig:FB_Freqs}
\end{figure*}

\begin{figure*}
\centering
\includegraphics[scale=0.5]{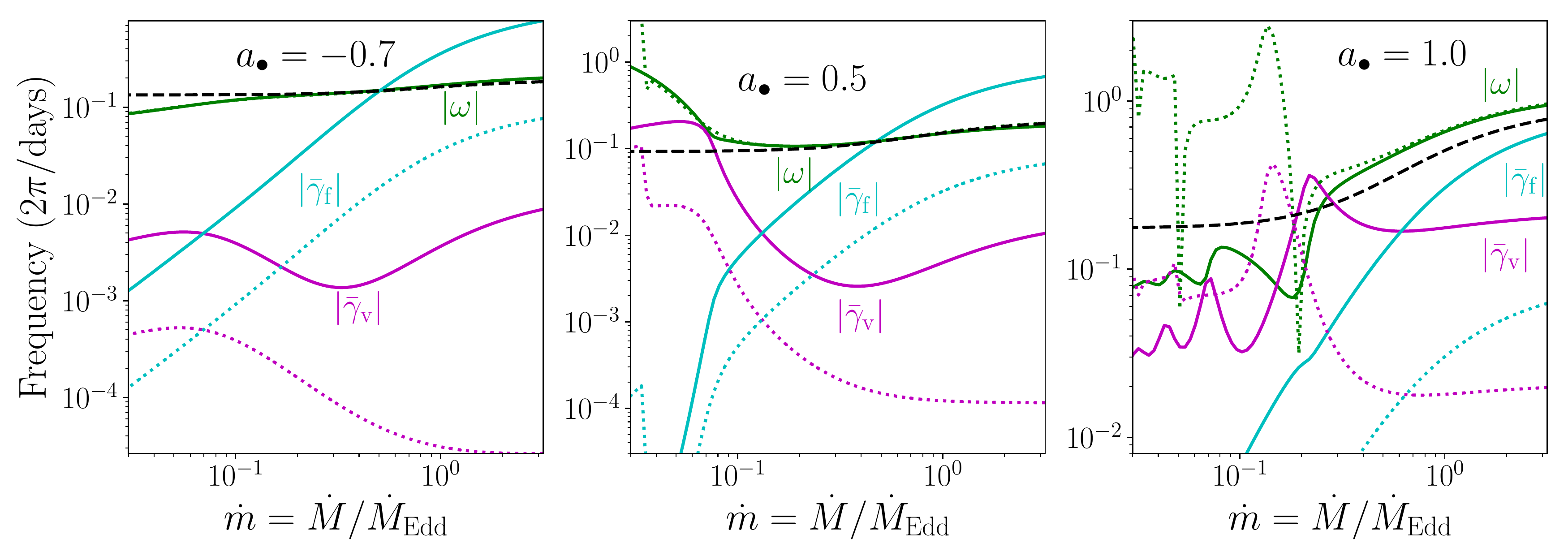}
\caption{ Same as Figure~\ref{fig:FB_Freqs}, except $\Mh = 10^7 \, \Msun$.}
\label{fig:FB_Freqs_largeMh}
\end{figure*}

Figure~\ref{fig:FB_Freqs} plots the precession frequency $\om$ and viscous [$\bcgv$, Eq.~\eqref{eq:bcgv}] and fall-back [$\bcgf$, Eq.~\eqref{eq:bcgf}] damping rates of the TDE disk.  We see the addition of the fall-back torque does little to modify $\om$ and $\bcgv$ of the disk (compare Fig.~\ref{fig:FB_Freqs} to Fig.~\ref{fig:TDE_Freqs}).  However, the fall-back torque can have a substantial impact on the disk's dynamical evolution, especially when the disk's accretion rate is high ($\dotm \gtrsim 1$).  For all SMBH spin parameters $\ah$ considered, $\bcgf$ exceeds $\om$ when the disk viscosity is high ($\ag = 0.1$), especially for low SMBH spins.  Even when the viscosity is low ($\ag = 0.01$), $\bcgf$ can exceed $\om$ when the SMBH spin is low ($\ah = 0.4$).  The damping rate $\bcgf$ becomes less than $\bcgv$ only for low accretion rates ($\dotm \lesssim 0.3-1.0$).

Figure~\ref{fig:FB_Freqs_largeMh} is identical to Figure~\ref{fig:FB_Freqs}, except the SMBH mass is larger ($\Mh = 10^7 \, \Msun$).
The qualitative dependence of the precession/damping rates $\om$, $\bcgf$, and $\bcgv$ on the disk and SMBH parameters ($\dotm$, $\ag$, and $\ah$) is similar.  The main difference between Figures~\ref{fig:FB_Freqs} and~\ref{fig:FB_Freqs_largeMh} is that the accretion rate below which the viscous damping rate dominates the fall-back damping rate ($|\bcgv| \gtrsim |\bcgf|$) is lower ($\dotm \lesssim 0.07-0.6$ for Fig.~\ref{fig:FB_Freqs_largeMh}, $\dotm \lesssim 0.3-1.0$ for Fig.~\ref{fig:FB_Freqs}).

The inclusion of the fall-back torque does not introduce any new features into the dynamical disk warp eigenfunctions $\tW$, except for a small ``kink'' at the disk's outer truncation radius $\rout$ ($\der \tbg/\der r|_{r=\rout} \ne 0$ and $\der \tphi/\der r|_{r=\rout} \ne 0$).  Since the fall-back torque has a negligible impact on the disk's warp profile, there are only small differences between the precession frequency $\om$ and viscous damping rate $\bcgv$ between Figures~\ref{fig:FB_Freqs} and~\ref{fig:TDE_Freqs} and Figures~\ref{fig:FB_Freqs_largeMh} and~\ref{fig:TDE_Freqs_largeMh}.

\subsection{Steady-State Warp Profiles}
\label{sec:FB_steady}

\begin{figure*}
\centering
\includegraphics[scale=0.7]{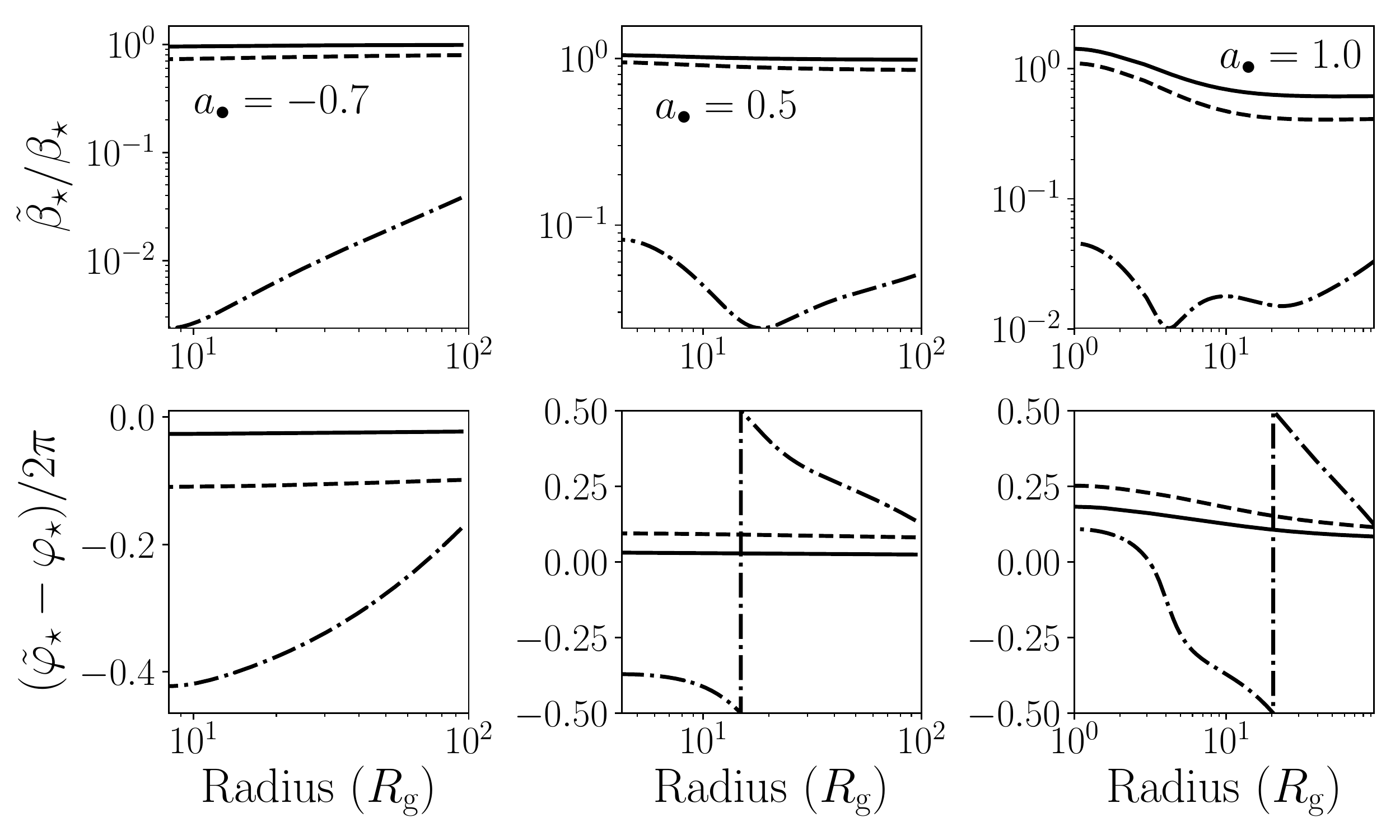}
\caption{Disk warp $\tbg_\star(r)$ (top panels) and twist $\tphi_\star(r)$ (bottom panels) radial profiles for the steady-state solution $\tWs = \tbg_\star e^{\im \tphi_\star}$, for $\dotm = 10$ (solid), $\dotm = 1$ (dashed), and $\dotm = 0.1$ (dot-dashed), with dimensionless SMBH spin parameters $\ah$ as indicated.  We normalize the solutions to the complex orbital angular momenta of the tidally disrupted star $W_\star = \beta_\star e^{\im \varphi_\star}$.  Here, $\ag = 0.1$, $\Mh = 10^6 \, \Msun$, $\Ms = 1 \, \Msun$, and $\Rs = 1 \, \Rsun$.  We assume $\dot M = \dotMfb$.}
\label{fig:SS_Warp_highv}
\end{figure*}

\begin{figure*}
\centering
\includegraphics[scale=0.7]{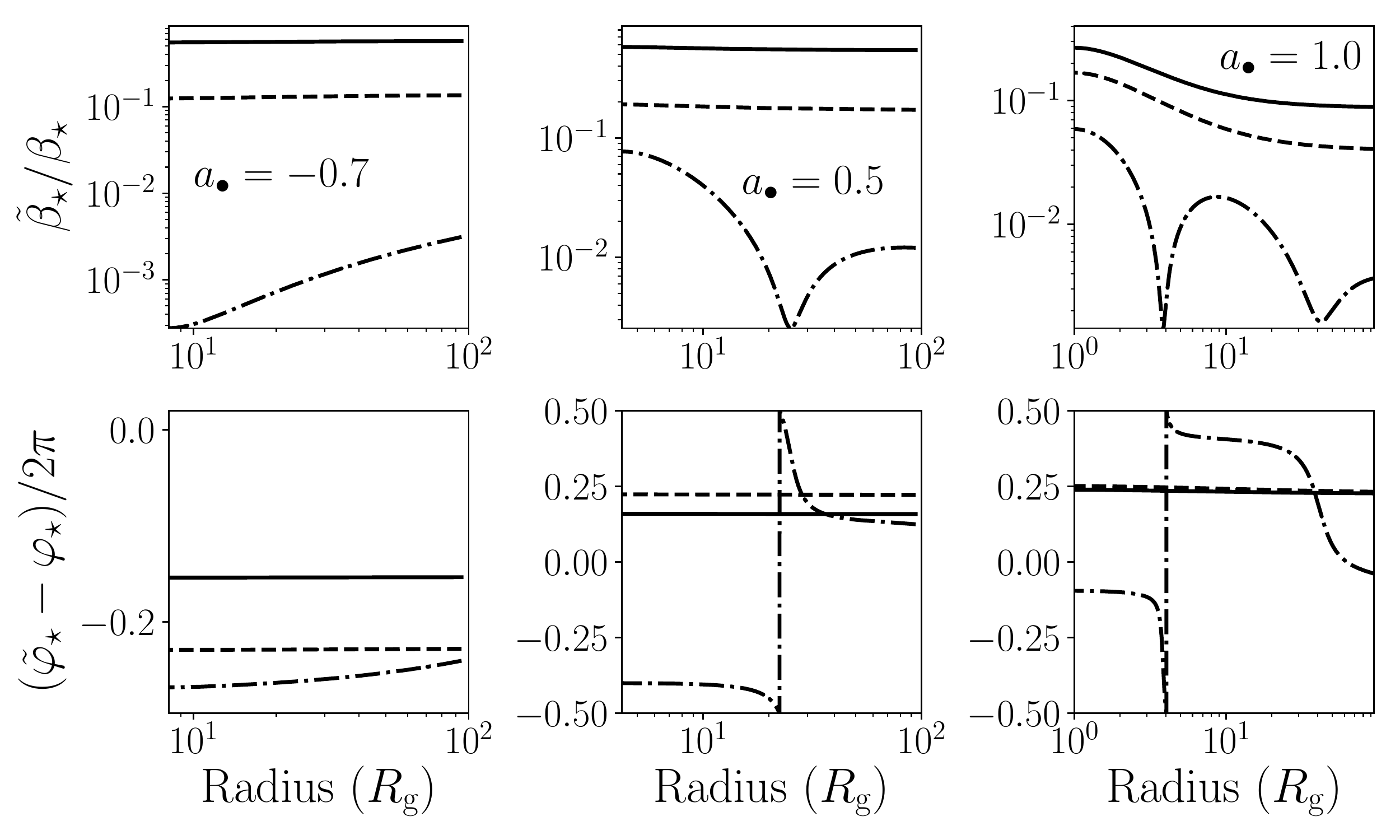}
\caption{Same as Figure~\ref{fig:SS_Warp_highv}, except $\ag = 0.01$.}
\label{fig:SS_Warp_lowv}
\end{figure*}

Figures~\ref{fig:SS_Warp_highv} and~\ref{fig:SS_Warp_lowv} plot the disk's complex steady-state warp profile $\tWs = \tbg_\star e^{\im \tphi_\star}$ for select accretion rate values $\dotm$ and SMBH spin parameters $\ah$ as indicated.  In many ways, the steady-state warp profiles are similar to their precessing warp profile counterparts ($\tW = \tbg e^{\im \tphi}$).  When the accretion rate is high ($\dotm \gtrsim 1$), only prograde and nearly-extremal SMBHs ($\ah \approx 1$) have warp profiles $\tbg_\star$ which varies substantially across the disk.  At these high $\dotm$ rates, the variation in the disk's twist $\tphi_\star$ is negligible, unless the SMBH is prograde, near extremal, and disk viscosity high ($\ag = 0.1$).  When the accretion rate is low ($\dotm \lesssim 1$), $\tbg_\star$ becomes non-trivial.  Low viscosity disks ($\ag = 0.01$) with prograde spins have more oscillatory $\tbg_\star$ in comparison to their high viscosity ($\ag = 0.1$) counterparts.  Retrograde disks have inner disk tilts nearly aligned with the SMBH's equitorial plane [$\tbg(\rin) \approx 0$] when the accretion rate is low ($\dotm \sim 0.1$).  High viscosity disks with low accretion rates have $\tphi_\star$ which increase or decrease steadily across the disk's radial extent (depending on the sign of $\ah$), while low viscosity $\tphi_\star$ variations are negligible unless near a warp node [when $\tbg_\star(r) \simeq 0$].

The main difference between the steady-state warp profiles $\tW_\star = \tbg_\star e^{\im \tphi_\star}$ (Figs.~\ref{fig:SS_Warp_highv}-\ref{fig:SS_Warp_lowv}) and the precession profiles $\tW = \tbg e^{\im \tphi}$ (Figs.~\ref{fig:TDE_Warp_highv}-\ref{fig:TDE_Warp_lowv}) are the normalization conditions, which cause the profiles to evolve differently as the disk's accretion rate $\dotm$ drops.  The $\tW_\star$ normalization cannot be freely chosen, and is determined by the tidally-disrupted star's orbital angular momentum $W_\star = \bg_\star e^{\im \varphi_\star}$.  As $\dotm$ decreases, so does the fall-back torque's magnitude, and it becomes more difficult to tilt the outer disk in opposition to the SMBH's Lense-Thirring torque.  As a result, $\tbg_\star$ decreases in magnitude across the entire disk when $\dotm$ is lowered.

\section{Discussion}
\label{sec:Discuss}

\subsection{Theoretical Uncertainties}
\label{sec:ThryUncert}

A major uncertainty in our work is how efficiently the fall-back material influences the disk warp [Eq.~\eqref{eq:vTf}].  We have adopted a simple prescription, and fixed the location of the fall-back angular momentum deposition to be at the outer truncation radius of the disk $\rout$ [Eq.~\eqref{eq:rout}].  Letting the angular momentum be deposited at locations $r_{\rm dep} \sim \rout$ will change the damping rates $\bcgf$ in Figures~\ref{fig:FB_Freqs}-\ref{fig:FB_Freqs_largeMh} by factors of order unity (see also \citealt{ShenMatzner(2014),XiangGruess(2016),Ivanov(2018)}).  Another uncertainty is our assumption $\dot M = \dotMfb$ when computing our damping rates in Figures~\ref{fig:FB_Freqs}-\ref{fig:FB_Freqs_largeMh}.  As the disk cools, the viscous time $t_{\rm v}$ [Eq.~\eqref{eq:tv}] will become longer than the timescale over which the fall-back torque decreases [$\Ms/2\dotMfb \sim \tf$, Eq.~\eqref{eq:tf}].  However, the fall-back accretion rate at these times is typically small, and fall-back damping $\bcgf$ will be negligible compared to viscous damping $\bcgv$.  If the disk is Eddington limited at early times ($\dot M \lesssim \dotMedd$) as suggested by some (e.g. \citealt{MetzgerStone(2016)}), then $\dot M \approx \dotMfb \approx \dotMedd = \text{constant}$ during the Eddington limited phase, and our results remain valid.

There are both theoretical justification and observational evidence that TDE disks may be eccentric.  Depending on the pericenter distance of the star's orbit and the SMBH spin, the eccentric debris streams can take anywhere from $t \sim 1 - 10 \ \tf$ to completely circularize, with long circularization times ($\sim 4-10 \ \tf$) the most common (e.g. \citealt{Rees(1988),Hayasaki(2013),Hayasaki(2016),Guillochon(2014),Piran(2015),GuillochonRamirez-Ruiz(2015),Shiokawa(2015),Bonnerot(2016),Krolik(2016)}).  Moreover, some emission lines from TDE debris are fit much better by modeling the accretion disk with an order-unity eccentricity \citep{Liu(2017),Cao(2018)}.  In order to understand how such eccentric disks are twisted under the competing influences of relativistic apsidal precession and internal pressure torques, the non-linear eccentric disk theory of \cite{Ogilvie(2001)} must be used (see also \citealt{OgilvieLynch(2019)}), and will also affect the energy dissipation rate in the disk \citep{BarkerOgilvie(2014),Chan(2018),WienkersOgilvie(2018)}.  A theory of eccentric and warped disks has yet to be developed, making it unclear how relaxing our assumption of circular disks will effect our results.

To model the warped TDE accretion disk, we have used fully relativistic expressions for the apsidal and nodal precession frequencies around the spinning SMBH (e.g. \citealt{Kato(1990)}), but neglected changes in length scales due to relativity (affecting $\pd/\pd r$) and time dilation (affecting $\pd/\pd t$) in the SMBH's accretion disk.  Although relativistic theories of warped accretion disks around spinning BHs have been developed \citep{IvanovIllarionov(1997),DemianskiIvanov(1997),ZhuravlevIvanov(2011)}, we chose to use the formalism  of \cite{Ogilvie(1999)} and \cite{LubowOgilvie(2000)} with the fully relativistic apsidal and nodal precession frequencies for simplicity.  We note these other warped disk theories are not fully relativistic: \cite{IvanovIllarionov(1997),DemianskiIvanov(1997)} include only the leading order post-Newtonian corrections to changes in length scales and time dilation, while \cite{ZhuravlevIvanov(2011)} assume a slowly spinning BH ($|\ah| \ll 1$).  Since the ISCO radius [Eq.~\eqref{eq:rin}] is approximatly equal to the BH's event horizon when the BH is prograde and near extremal ($\ah \approx 1$), order unity BH spins must be included in a relativistic theory to fully understand how time dilation modifies the TDE disk's precession frequency.

\subsection{Observational Implications}
\label{sec:ObsImps}

When the accretion rate is high ($\dot M \gtrsim 0.4 \, \dotMedd$), we have shown the rigid-body Lense-Thirring precession frequency $\bomhr$ [Eq.~\eqref{eq:bomh_rigid}] is a good approximation to the lowest-order precession frequency $\om$ of the disk (Figs.~\ref{fig:TDE_Freqs}-\ref{fig:TDE_Freqs_largeMh} \&~\ref{fig:FB_Freqs}-\ref{fig:FB_Freqs_largeMh}), differing by factors of $\sim 2-3$ only when the SMBH spin is prograde and sufficiently high ($\ah \gtrsim 0.9$, Fig.~\ref{fig:TDE_Freqs_ah}).  But when the TDE disk's accretion rate is low ($\dot M \lesssim 0.4 \, \dotMedd$), the disk's precession frequency $\om$ can differ from $\bomhr$ substantially.  Figures~\ref{fig:TDE_Freqs}-\ref{fig:TDE_Freqs_largeMh} and~\ref{fig:FB_Freqs}-\ref{fig:FB_Freqs_largeMh} show the deviation of $\om$ from $\bomhr$ can be a factor of a few for disks with high viscosities ($\ag = 0.1$) or with retrograde SMBH spins ($\ah < 0$), and be more than an order of magnitude for disks with low viscosities ($\ag = 0.01$) and prograde SMBH spins ($\ah > 0$).  The precession frequency of TDE disks with prograde SMBH spins, low accretion rates, and low viscosities can vary with $\dot M$ in such a dramatic manner, that it is nearly impossible to get any information on the SMBH by analyzing the TDE disk's precession rate (see Sec.~\ref{sec:LT_freqs} for discussion of the reason behind a highly variable $\om$ with $\dot M$).  Any stable detected quasi-periodic oscillations in TDEs from precessing accretion disks (e.g. \citealt{Burrows(2011),Saxton(2012b)}) must therefore be either in a high accretion phase ($\dot M \gtrsim 0.4 \, \dotMedd$), have high viscosities ($\ag \sim 0.1$), or be orbiting retrograde around the SMBH's spin.

 When the fall-back material has a negligible influence on the precessing disk's evolution, we find a wide range of viscous alignment timescale of the TDE disk with the SMBH's equitorial plane.  The viscous alignment timescales range anywhere between a few years to a few days, depending on the disk's accretion rate and viscosity (Figs.~\ref{fig:TDE_Freqs}-\ref{fig:TDE_Freqs_largeMh}), as well as the SMBH spin (Fig.~\ref{fig:TDE_Freqs_ah}).  The viscous damping rate is typically at least an order of magnitude below the precession frequency, unless the SMBH spin is prograde and high ($\ah \gtrsim 0.9$) and viscosity parameter is large ($\ag \sim 0.1$).  The TDE disk should therefore stably precess around the SMBH spin vector for many precession periods unless the SMBH is prograde and near extremal, with a high disk viscosity.  When the accretion rate is sufficiently low ($\dot M \sim 0.1 \, \dotMedd$) and SMBH spinning prograde, the viscous damping rates $\bcgv$ are comparable to the precession frequencies $\om$ for high viscosity disks ($\ag = 0.1$), and can exceed the $\om$ values of low viscosity disks ($\ag = 0.01$).
 In contrast, viscous damping rates for disks around retrograde SMBH spins have viscous damping rates orders of magnitude below the disk's precession frequency, regardless of the accretion rate (for the accretion rate values investigated).
  Therefore, when the accretion rate is low, coherent precession is unlikely to be detectable due to the rapid alignment of the TDE disk with the SMBH's equatorial plane, unless the disk orbits retrograde with respect to the SMBH's spin.

In contrast, the fall-back alignment rates are typically comparable to or exceed the disk's precession frequency when the disk's accretion rate is high ($\dot M \gtrsim \dotMedd$; Figs.~\ref{fig:FB_Freqs}-\ref{fig:FB_Freqs_largeMh}).  The inclusion of the fall-back torque causes the initially misaligned and precessing TDE disk to rapidly evolve into its steady-state warp profile.  For the TDE parameters investigated in this work, the disk warp evolves to its steady-state profile in a few to tens of days.  If quasi-periodic oscillations in the hard X-ray from tidal disruption flares are emitted by a precessing accretion disk (e.g. \citealt{Burrows(2011),Saxton(2012b)}), then the fall-back material must deposit far less angular momentum to the accretion disk than we assumed with our prescription~\eqref{eq:vTf}.

For both the rigidly precessing (Figs.~\ref{fig:TDE_Warp_highv}-\ref{fig:TDE_Warp_lowv}) and steady-state (Figs.~\ref{fig:SS_Warp_highv}-\ref{fig:SS_Warp_lowv}) warp profiles of the TDE disk, the inner disk has a \textit{higher} tilt to the SMBH's equitorial plane than the outer disk [$\tbg(\rin) > \tbg(\rout)$, $\tbg_\star(\rin) > \tbg_\star(\rout)$] for prograde SMBH spins, in sharp contrast to the ``standard'' picture dating back to \cite{BardeenPetterson(1975)}.
 The inner edge of the disk is less tilted than the outer edge when the SMBH spin is retrograde.
  Previous models of warped TDE disks obtained different results because they neglected the dominant internal torque (pressure rather than viscosity) acting the disk (e.g. \citealt{Lei(2013)}).  Including the highly tilted inner edge of a TDE accretion disk with a prograde SMBH spin will further constrain models explaining the variability in the hard X-ray of jetted TDEs with Lense-Thirring precession \citep{StoneLoeb(2012)}, since the TDE jet is likely to be tightly coupled to the inner edge of the accretion disk \citep{Liska(2018)}.

\section{Conclusions}
\label{sec:Conc}

We have carried out a systematic analysis of the dynamics and evolution of warped accretion disks that are misaligned with the equatorial plane of the central SMBH.  Such disks are naturally produced in TDEs when the stellar orbital angular momentum axis is misaligned with the BH spin axis.  Even with our somewhat idealized model of the TDE disks, our work clarifies a number of disagreements in the literature, and uncovers several new dynamical behaviors of the TDE disk evolution.
 
Section~\ref{sec:DiskWarp_LT} examines the warp profile, precession and viscous damping rates of simple disk models (powerlaw $\Sg$, constant $H/r$) around a BH.  We find that to properly calculate the warp profile, it is important to include pressure torques, which dominate viscous torques because of relativistic apsidal precession.  The inner disk is generally more (less) tilted than the outer disk for prograde (retrograde) BH spins (Figs.~\ref{fig:Warp_LT_ah=0.5}-\ref{fig:Warp_LT_ah=1.0}).  The global disk precession frequency and viscous damping rate can vary by more than an order of magnitude as the disk scaleheight is varied (Fig.~\ref{fig:ToyFreqs_H/r}), due to the sensitive dependence of the effective warp potential on the disk scaleheight [Eq.~\eqref{eq:V_eff}].

Section~\ref{sec:TDEModel} constructs a simple model for the TDE disk soon after the star tidally disrupts.  We obtain analytic prescriptions for the disk's surface density and aspect ratio (Fig.~\ref{fig:Disk_Profile}), which depend on the accretion rate.

Section~\ref{sec:TDEDisk_LT} uses our analytic disk model (Sec.~\ref{sec:TDEModel}) to calculate a TDE disk's tilt profile, as well as the precession and damping rates of the disk with respect to the SMBH's equatorial plane.  Like the simple disk model studied in Section~\ref{sec:DiskWarp_LT}, we find the inner disk to be more (less) tilted than the outer disk for prograde (retrograde) SMBH spins, and the tilt profile to become more oscillatory at lower accretion rates and viscosities (Figs.~\ref{fig:TDE_Warp_highv}-\ref{fig:TDE_Warp_lowv}).  Disks with high accretion rates have global precession frequencies which closely match the disk's rigid-body Lens-Thirring precession frequency, but disks with low accretion rates and prograde SMBH spins can have their precession frequencies differ from the rigid-body frequency by an order of magnitude (Figs.~\ref{fig:TDE_Freqs}-\ref{fig:TDE_Freqs_largeMh}).
 
Section~\ref{sec:TDEDisk_FB} examines how angular momentum deposition by fall-back material affects the warp structure and inclination evolution of the TDE disk.  The main effect of the fall-back material is to cause the disk tilt to rapidly evolve to its steady-state profile, over a timescale shorter than the disk's global precession period (Figs.~\ref{fig:FB_Freqs}-\ref{fig:FB_Freqs_largeMh}).  The steady-state warp profiles have a similar structure as the precessing warp profiles, except the steady-state warp amplitude decreases with the disk's accretion rate (Figs.~\ref{fig:SS_Warp_highv}-\ref{fig:SS_Warp_lowv}).

\section*{Acknowledgments}
We thank the referee, Pavel Ivanov, for many helpful comments and suggestions which significantly improved the quality of this work.  JZ thanks Gordon Ogilvie, Rong-Feng Shen, and Almog Yalinewich for helpful discussions.  This work is supported in part by the NSF grant AST-
1715246 and NASA grant NNX14AP31G.  JZ was supported in part by a graduate NASA Earth and Space Sciences Fellowship in Astrophysics.

\appendix

\section{Density Wave Dispersion Relation in a Viscous, Non-Keplerian Disk}
\label{sec:DispRel}

As discussed in Section~\ref{sec:DiskWarp_LT}, the behavior of bending waves depends critically on the disk aspect ratio $H/r$ in comparison to the Shakura-Sunyaev viscosity parameter $\ag$ and dimensionless non-Keplerian epicyclic frequency $\tkg = (\Om^2-\kg^2)/2\Om^2$.  This section shows this condition may be understood using WKB theory for inertial-density waves.

Consider an accretion disk with vertically isothermal sound-speed $\cs = H \Om$, unperturbed density $\rg(r,z) = \rg(r) e^{-z^2/2H^2}$, pressure $p(r,z) = \cs^2(r) \rho(r,z)$, and azimuthal fluid velocity $v_\varphi(r) = r \Om$, and with radial $v_r$ and vertical $v_z$ velocity components equal to zero.  We perturb each equilibrium state quantity $X$ by a perturbation $\dg X$ which satisfies
\be
\frac{\pd \dg X}{\pd r}, \frac{\pd \dg X}{\pd z} \gg \frac{1}{r} \frac{\pd \dg X}{\pd \varphi} \sim \frac{\dg X}{r}.
\ee
Moreover, we assume the disk is thin ($H/r \ll 1$), so the equilibrium quantities $X(r,z)$ satisfy
\be
\frac{\pd X}{\pd z} \gg \frac{\pd X}{\pd r} \sim \frac{X}{r}.
\ee
Decomposing the azimuthal and time dependences of the perturbations $\dg X$ as $\dg X(r,z,\varphi,t) = \dg X(r,z) e^{\im (m \varphi - \om t)}$, we have (e.g. \citealt{FuLai(2009)})
\begin{align}
&- \im \pom \dg \rho + \rho \frac{\pd}{\pd r} \dg v_r + \frac{\pd}{\pd z} (\rho \dg v_z) = 0,
\label{eq:drgdt} \\
&-\im \pom \dg v_r - 2 \Om \dg v_\vphi = -\frac{1}{\rho} \frac{\pd}{\pd r} \dg p + (\bfv)_r,
\label{eq:dvrdt} \\
&-\im \pom \dg v_\vphi + \frac{\kg^2}{2 \Om} \dg v_r = (\bfv)_\vphi,
\label{eq:dvphidt} \\
&-\im \pom \dg v_z = - \frac{1}{\rg} \frac{\pd}{\pd z} \dg p + \frac{1}{\rho^2} \frac{\der p}{\der z} \dg \rg + (\bfv)_z,
\label{eq:dvzdt}
\end{align}
where
\begin{align}
(\bfv)_r & = \nu \left( \frac{4}{3} \frac{\pd^2}{\pd r^2} + \frac{\pd \ln \rho}{\pd z} \frac{\pd}{\pd z}+ \frac{\pd^2}{\pd z^2} \right) \dg v_r
\nonumber  \\
&+ \nu \left( - \frac{2}{3} \frac{\pd^2}{\pd r \pd z} + \frac{\pd \ln \rg}{\pd z} \frac{\pd}{\pd r} + \frac{\pd^2}{\pd r \pd z} \right) \dg v_z,
\label{eq:fvr} \\
(\bfv)_\vphi &= \nu \left( \frac{\pd^2}{\pd r^2} + \frac{\pd \ln \rg}{\pd z} \frac{\pd}{\pd z} + \frac{\pd^2}{\pd z^2} \right) \dg v_\vphi,
\label{eq:fvphi} \\
(\bfv)_z &= \nu \left( - \frac{2}{3} \frac{\pd \ln \rg}{\pd z} \frac{\pd}{\pd r} + \frac{1}{3} \frac{\pd^2}{\pd r \pd z} \right) \dg v_r
\nonumber \\
&+ \nu \left( \frac{\pd^2}{\pd r^2} + \frac{4}{3} \frac{\pd \ln \rho}{\pd z} \frac{\pd}{\pd z} + \frac{4}{3} \frac{\pd^2}{\pd z^2} \right) \dg v_z,
\label{eq:fvz}
\end{align}
are the viscous force terms, and
\be
\pom = \om - m \Om.
\ee
We assume $m = \cO(1)$ throughout this section.

\subsection{High Vertical Wavenumber Limit}

When the vertical gradients of the fluid perturbations satisfy $\pd \dg X/\pd z \gg \dg X/H$, we may assume $\dg X \propto e^{\im (\kr r + \kz z)}$, and equations~\eqref{eq:drgdt}-\eqref{eq:dvzdt} become
\begin{align}
&-\im \pom \dg \bar \rg + \im \kr r \Om \dg \bar v_r + \im \kz r \Om \dg \bar v_z = 0,
\label{eq:drgdt_highkz} \\
&-\im \pom_r r \Om \dg \bar v_r - 2 r \Om^2 \dg \bar v_\vphi = -\im \kr \cs^2 \dg \bar \rho - \ag r \cs^2 k_{rz}^2 \dg \bar v_z,
\label{eq:dvrdt_highkz} \\
&-\im \pom_\vphi r \Om \dg \bar v_\vphi + \frac{r \kg^2}{2} \dg \bar v_r = 0
\label{eq:dvphidt_highkz} \\
&-\im \pom_z r \Om \dg \bar v_z = -\im \cs^2 k_z \dg \bar \rg - \ag r \cs^2 k_{zr}^2 \dg \bar v_r,
\label{eq:dvzdt_highkz}
\end{align}
 where $\dg \bar \rho = \dg \rho/\rho$, $\dg \bar \bv = \dg \bv/r \Om$,
\begin{align}
k_{rr}^2 &= \frac{4}{3} \kr^2 + \kz^2, \\
k_{\vphi \vphi}^2 &= k_r^2 + \kz^2, \\
k_{zz}^2 &= \kr^2 + \frac{4}{3} \kz^2, \\
k_{rz}^2 &= -\frac{2}{3} \kr \kz + \kz \kr = \frac{1}{3} \kr \kz, \\
k_{zr}^2 &= - \frac{2}{3} \kz \kr + \kr \kz = \frac{1}{3} \kr \kz,
\end{align}
and
\begin{align}
\pom_r &= \pom + \im \ag r \cs^2 k_{rr}^2, \\
\pom_\vphi &= \pom + \im \ag r \cs^2 k_{\vphi \vphi}^2, \\
\pom_z &= \pom + \im \ag r \cs^2 k_{zz}^2.
\end{align}

Equations~\eqref{eq:drgdt_highkz}-\eqref{eq:dvzdt_highkz} may be solved for the dispersion relation
\begin{align}
&(\pom_r \pom_\vphi - \kg^2)(\Om^2 - \kz^2 \cs^2)\Om^2 
\nonumber \\
&+ \ag \cs^4 \pom_\vphi \big[ \ag \pom k_{rz}^2 k_{zr}^2 + \im \Om \kr \kz (k_{rz}^2 + k_{zr}^2) \big]
\nonumber \\
&= \kr^2 \cs^2 \pom_z \pom_\vphi \Om^2.
\label{eq:disp_highkz}
\end{align}
From Equation~\eqref{eq:disp_highkz}, we see the cross terms $k_{rz}$ and $k_{zr}$ are negligible when $\ag \kz^2 \cs^2 \ll \Om^2$.  Assuming $\kz = \sqrt{n}/H$, where $n$ the order of the wave and measures the number of vertical nodes, we see the cross terms are negligible when $n \ag \ll 1$.  The next section derives the dispersion relation for these low-order density waves [$n = \cO(1)$].

\subsection{Low-Order Inertial-Density Waves}

This section derives the dispersion relation for low-wavenumber density waves ($n \ll \ag^{-1}$).  Using the fact that the cross viscous force terms are negligible when $n$ is sufficiently low, the viscous force components reduce to
\begin{align}
(\bfv)_r &\simeq \ag H^2 \Om \left( \frac{4}{3} \frac{\pd^2}{\pd r^2} + \frac{\pd \ln \rg}{\pd z} \frac{\pd}{\pd z} + \frac{\pd^2}{\pd z^2} \right) \dg v_r,
\label{eq:fvr_lowkz} \\
(\bfv)_\vphi &\simeq \ag H^2 \Om \left( \frac{\pd^2}{\pd r^2} + \frac{\pd \ln \rho}{\pd z} \frac{\pd}{\pd z} + \frac{\pd^2}{\pd z^2} \right) \dg v_\vphi,
\label{eq:fvphi_lowkz} \\
(\bfv)_z &\simeq \ag H^2 \Om \left( \frac{\pd^2}{\pd r^2} + \frac{4}{3} \frac{\pd \ln \rg}{\pd z} \frac{\pd}{\pd z} + \frac{4}{3}\frac{\pd^2}{\pd z^2} \right) \dg v_z.
\label{eq:fvz_lowkz}
\end{align}
Since $\pd \ln \rg/\pd z = -z/H^2$, one may decompose the vertical dependence of the fluid perturations in terms of Hankel Functions $H_n(Z)$:
\be
H_n(Z) = (-1)^n e^{Z^2/2} \left( \frac{\der}{\der Z} \right)^n e^{-Z^2/2}.
\ee
Specifically, writing
\begin{align}
\dg \rho &= \rho \dg \bar \rg H_n \left( \frac{z}{H} \right) e^{\im \kr r}, \\
\dg v_r &= r \Om \dg \bar v_r H_n \left( \frac{z}{H} \right) e^{\im \kr r}, \\
\dg v_\vphi &= r \Om \dg \bar v_\vphi H_n \left( \frac{z}{H} \right) e^{\im \kr r}, \\
\dg v_z &= r \Om \dg \bar v_z H_n' \left( \frac{z}{H} \right) e^{\im \kr r},
\end{align}
where $H_n'(Z) = \der H_n/\der Z$, the fluid perturbation equations reduce to
\begin{align}
&-\im r \Om \pom \dg \bar \rg + \im \kr r^2 \Om^2 \dg \bar v_r - k_z^2 \cs r^2 \Om \dg v_z = 0,
\label{eq:drgdt_lowkz} \\
&-\im r \Om \pom_r \dg \bar v_r - 2 r \Om^2 \dg \bar v_\vphi + \im \cs^2 \kr \dg \bar \rho = 0,
\label{eq:dvrdt_lowkz} \\
&-\im r \Om \pom_\vphi \dg \bar v_\vphi + \frac{r \kg^2}{2} \dg \bar v_r = 0,
\label{eq:dvphidt_lowkz} \\
&-\im r \Om \pom_z \dg \bar v_z + \cs \Om \dg \bar \rg = 0,
\label{eq:dvzdt_lowkz}
\end{align}
where $\kz = \sqrt{n}/H$ and
\begin{align}
\pom_r &= \pom + \im \ag H^2 \Om \bigg( \frac{4}{3} \kr^2 + \kz^2 \bigg) \\
\pom_\vphi &= \pom + \im \ag H^2 \Om \bigg( \kr^2 + \kz^2 \bigg) \\
\pom_z &= \pom + \im \ag H^2 \Om \bigg[ \kr^2 + \frac{4(n-1)}{3n} \kz^2 \bigg].
\end{align}
Equations~\eqref{eq:drgdt_lowkz}-\eqref{eq:dvzdt_lowkz} yield the dispersion relation
\be
(\pom \pom_z - n \Om^2)(\pom_r \pom_\vphi - \kg^2) = \cs^2 \kr^2 \pom_r \pom_\vphi.
\label{eq:disp_lowkz}
\ee

\subsection{Limiting Case: Bending Waves}

For a bending wave ($m = n = 1$), dispersion relation~\eqref{eq:disp_lowkz} for a low-frequency ($\om \ll \Om$), long radial wavelength ($\kr \ll H^{-1}$) disk with $\ag \ll 1$ reduces to 
\be
\om^2 + (\im \ag + \tkg) \Om \om - \frac{1}{4} \kr^2 \cs^2 \simeq 0.
\label{eq:disp_BW}
\ee 
Equation~\eqref{eq:disp_BW} may be solved for the group velocity of the bending wave $v_{\rm bw} = \der \om/\der \kr$:
\be
v_{\rm bw} = \pm \frac{\kr H \cs}{2 \sqrt{(\im \ag + \tkg)^2 + \kr^2 H^2}}.
\label{eq:v_bw}
\ee
The group velocity~\eqref{eq:v_bw} expresses the efficiency of angular momentum exchange by bending waves.

When $|\im \ag + \tkg| \lesssim \kr H$, Equation~\eqref{eq:v_bw} reduces to
\be
v_{\rm bw} \approx \pm \frac{1}{2} \cs.
\ee
In other words, when the disk viscosity parameter $\ag$ and dimensionless non-Keplerian epicyclic frequency $\tkg$ are sufficiently low, bending waves travel at half the sound-speed.  When $|\im \ag + \tkg| \gtrsim \kr H$, Equation~\eqref{eq:v_bw} reduces to a different expression:
\be
v_{\rm bw} \approx \pm \frac{\kr H \cs}{2(\im \ag + \tkg)}.
\ee
Thus, when $\ag$ becomes sufficiently large, the bending waves becomes diffusive, while when $|\tkg|$ becomes sufficiently large, bending waves travel at speeds significantly less than $\cs/2$.  For the long radial wavelength bending waves of interest for this work ($\kr \sim r^{-1}$), the condition for bending waves to travel at $|v_{\rm bw}| \gtrsim |\cs/2|$ is
\be
\ag \lesssim \frac{H}{r}
\hspace{5mm}
\text{and}
\hspace{5mm}
\tkg \lesssim \frac{H}{r}.
\ee

\section{Toy Model of a Black Hole Disk Warp}
\label{sec:LT_Toy}

\begin{figure}
\includegraphics[scale=0.65]{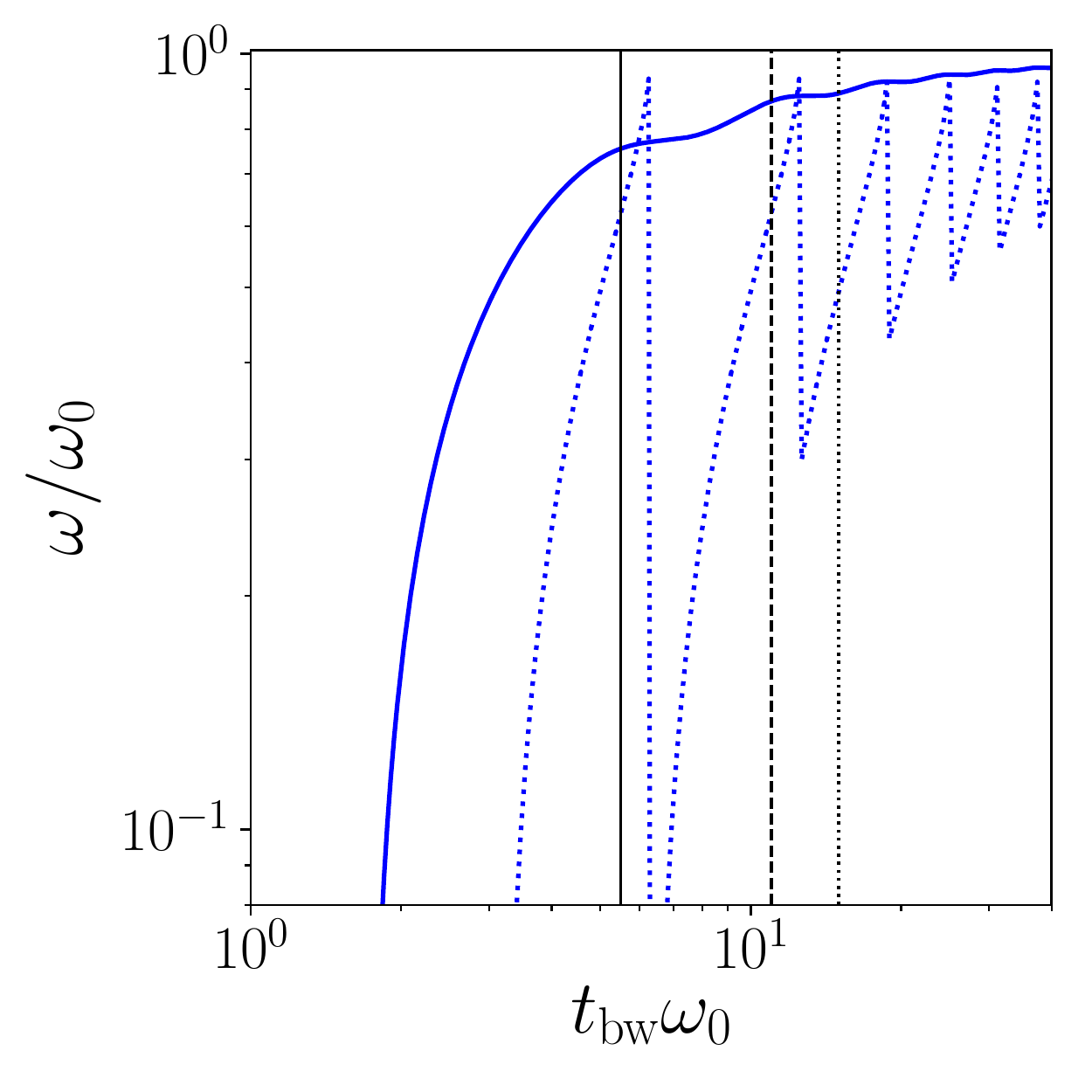}
\caption{Numerically [solid; Eq.~\eqref{eq:Warp}]  and analytically [dotted, Eq.~\eqref{eq:om_toy}] calculated eigenfrequencies $\om$ for our toy-model.  Vertical black lines denote $\tbw \om_0$ values of warp amplitude eigenfunctions $|W|$ displayed in Figure~\ref{fig:W_toy}.  Here, $\Sg \propto r^{-3/2}$ and $x_{\rm out} = 2$ ($r_{\rm out} = 7.39 \, \Rg$).}
\label{fig:om_toy}
\end{figure}

\begin{figure}
\includegraphics[scale=0.65]{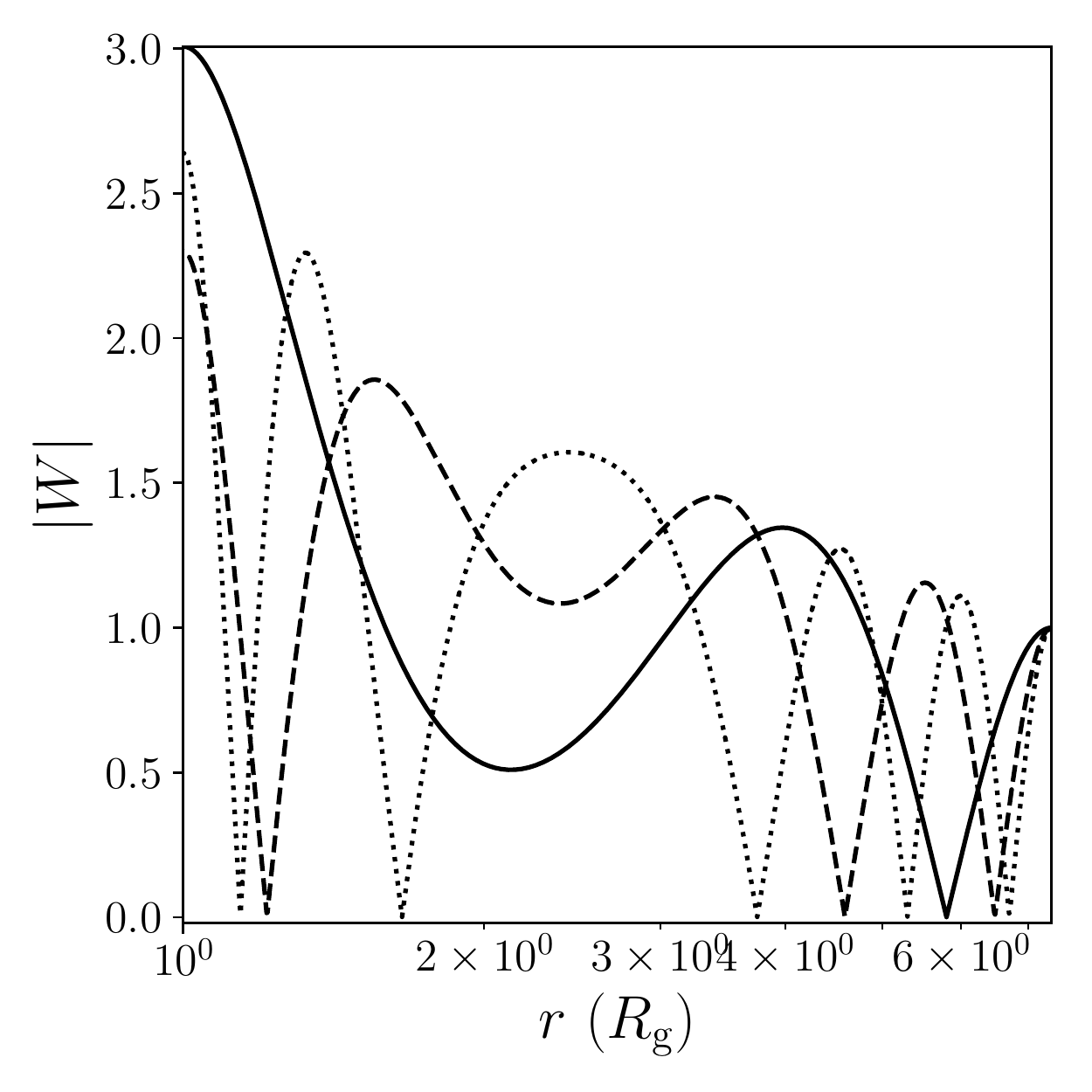}
\caption{Numerically calculated eigenfunctions [Eq.~\eqref{eq:Warp}] for the warp amplitude $|W|$, for $\tbw = 5.5$ (solid), $\tbw = 11$ (dashed), and $\tbw = 15$ (dotted).  See Fig.~\ref{fig:om_toy} for the eigenfrequencies $\om$.  Here, $\Sg \propto r^{-3/2}$ and $x_{\rm out} = 2$ ($r_{\rm out} = 7.39 \, \Rg$).}
\label{fig:W_toy}
\end{figure}

We begin with the warped disk equations [see Eqs.~\eqref{eq:dWdt} \&~\eqref{eq:G_res}]
\begin{align}
&\Sg r^2 \Om \frac{\pd W}{\pd t} = \frac{1}{r} \frac{\pd G}{\pd r} + \im \Sg r^2 \Om^2 \tOmp W,
\label{eq:dWdt_toy} \\
&\frac{\pd G}{\pd t} = \im \tkg \Om G + \frac{\Sg \cs^2 r^3 \Om}{4} \frac{\pd W}{\pd r},
\label{eq:dGdt_toy}
\end{align}
where $\tOmp = (\Om^2 - \Om_\perp^2)/2\Om^2$ is the dimensionless non-Keplerian nodal precession rate, $\Om_\perp$ is the nodal precession rate, and all other quantities are the same as usual.  We assume the disk is inviscid ($\ag = 0$).  Looking for eigenmode solutions of the form $W,G \propto e^{\im \om t}$, the warp equations may be re-arranged to give
\be
\frac{\pd}{\pd r} \left[ \frac{\Sg \cs^2 r^3 \Om}{4(\om - \tkg \Om)} \frac{\pd W}{\pd r} \right] + \Sg r^3 \Om (\om - \tOmp \Om)W = 0.
\label{eq:Warp}
\ee
When $\cs^2 \ll r^2 (\om - \tOmp \Om)(\om - \tkg \Om)$, we may assume $\pd^2 W/\pd r^2 \gg \pd W/\pd r$, and the above equation simplifies to
\be
\frac{\pd^2 W}{\pd r^2} + \frac{4(\om - \tOmp \Om)(\om - \tkg \Om)}{\cs^2} W \simeq 0.
\ee
For the rest of this section, we assume $\cs = \text{constant}$.  Letting $\tbw = 2 r/\cs$ and $x = \ln(r/\Rg)$, the above equation may be re-arranged to give
\be
\frac{\pd^2 W}{\pd x^2} + \tbw^2 (\om - \tOmp \Om)(\om - \tkg \Om)W = 0.
\label{eq:Warp_WKB}
\ee

Let
\be
V(x) = \tbw^2(\om - \tOmp \Om)(\om - \tkg \Om),
\ee
and defining
\be
k = \sqrt{|V|},
\ee
the WKB solution of Equation~\eqref{eq:Warp_WKB} is
\be
W(x) = \frac{A_+}{\sqrt{k}} e^{+\im \int^x k \der x'} + \frac{A_-}{\sqrt{k}} e^{-\im \int^x k \der x'}
\ee
when $V(x) > 0$, and
\be
W(x) = \frac{B_+}{\sqrt{k}} e^{+\int^x k \der x'} + \frac{B_-}{\sqrt{k}} e^{-\int^x k \der x'}
\ee
when $V(x) < 0$, where $A_\pm$, $B_\pm$ are constants to be determined by the boundary conditions.

For simplicity, we assume the disk is truncated at $x_{\rm in} = 0$ ($r_{\rm in} = \Rg$) and leave $x_{\rm out}$ ($r_{\rm out} = e^{x_{\rm out}}\Rg$) to be a free parameter.  We examine the disk's eigenmodes for a toy model:
\be
\tOmp = \tkg = x \om_0/\Om
\label{eq:toy_model}
\ee
We assume the usual torque-free boundary conditions:
\be
\left.\frac{\pd W}{\pd x}\right|_{x=0} = \left.\frac{\pd W}{\pd x}\right|_{x=x_{\rm out}} = 0,
\ee
with a normalization condition $W(x_{\rm out}) = 1$.  We look for low-frequency solutions ($|\om| \le \om_0$).  We define $x_c = \om/\om_0$ as the critical radius where $V(x_c) = 0$.  Notice with our toy model, $V(x) \ge 0$ everywhere.

The outer boundary condition gives [assuming $k(x_{\rm out}) \gg 1$]
\be
W(x) \simeq \sqrt{ \frac{k(x_{\rm out})}{k} } \cos \left( \int_x^{x_{\rm out}} k \der x' \right),
\ee
while the inner boundary condition is satisfied when [assuming $k(0) \gg 1$]
\be
\int_0^{x_{\rm out}} |x - x_c| \der x \simeq \frac{\pi n}{t_{\rm bw} \om_0}.
\ee
This equation has solutions
\be
\frac{\om}{\om_0} = \frac{x_{\rm out}}{2} \left( 1\pm \sqrt{ \frac{4\pi n}{x_{\rm out} \tbw \om_0} -1}\right),
\label{eq:om_n}
\ee
where $n$ is an integer.  Requiring the square root in expression~\label{eq:om_n} to be positive forces
\be
n > \frac{x_{\rm out}\tbw \om_0}{4\pi}.
\ee
Therefore, the lowest order eigenvalue $\om$ is given by
\be
\om = \frac{\om_0 x_{\rm out}}{2}  \left( 1 - \sqrt{ \frac{4\pi n_{\rm min}}{x_{\rm out}\tbw \om_0} -1}\right),
\label{eq:om_toy}
\ee
where
\be
n_{\rm min} = \left\lceil \frac{x_{\rm out} \tbw \om_0}{4\pi} \right\rceil,
\ee
and $\lceil \dots \rceil$ is the ceiling function.  Note that $n$ counts the number of nodes in the disk's warp amplitude [when $W(x) = 0$].

Figure~\ref{fig:om_toy} plots the numerically and analytically computed eigenfrequencies $\om$ as a function of $\tbw \om_0$.  Eigenvalues computed numerically solve Equation~\eqref{eq:Warp} using a shooting algorithm, while the analytic eigenvalues are given by Equation~\eqref{eq:om_toy}.  Although these eigenfrequencies quantitative values differ by a factor of $\sim \om_0$ due to the crudeness of the WKB approximation, both display oscillations in the rigid-body precession frequency $\om$ as $\tbw \om_0$ increases.  This is because when $\tbw \om_0$ increases, the $n_{\rm min}$ value of the disk's lowest-order eigenmode changes.  This causes an increase in the disk's eigenfrequency $\om$.  A characteristic of the disk's eigenfunction $W(x)$ when $n_{\rm min}$ changes values is the number of nodes [when $W(x) = 0$] increases.

The fact that the number of nodes changes at each of the eigenfrequency peaks is shown clearly in Figure~\ref{fig:W_toy}.  Displayed are the eigenfunctions for the $\tbw \om_0$ values marked by vertical black lines in Fig.~\ref{fig:om_toy}.  As $\tbw \om_0$ increases, so does the number of nodes in the disk's warp amplitude for the lowest-order eigenmode.

This toy model is analogous to a disk around a spinning black hole, since a black hole disk's effective potential $V(r) > 0$ over most of the disk's radial extent [see Eqs.~\eqref{eq:tkg}, \eqref{eq:omh}, \&~\eqref{eq:V_eff}].  The the precession frequency of a disk around a spinning black hole also has a sensitive and non-monotonic dependence on the bending wave crossing timescale $\tbw$, or equivalently the disk scaleheight $H/r$.


\begin{thebibliography}{99}

\bibitem[\protect\citeauthoryear{Abramowicz et al.}{1995}]{Abramowicz(1995)} Abramowicz M.~A., Chen X., Kato S., Lasota J.-P., Regev O., 1995, ApJ, 438, L37

\bibitem[\protect\citeauthoryear{Abramowicz et al.}{1988}]{Abramowicz(1988)} Abramowicz M.~A., Czerny B., Lasota J.~P., Szuszkiewicz E., 1988, ApJ, 332, 646

\bibitem[\protect\citeauthoryear{Arcavi et al.}{2014}]{Arcavi(2014)} Arcavi I., et al., 2014, ApJ, 793, 38

\bibitem[\protect\citeauthoryear{Bade, Komossa \& Dahlem}{1996}]{Bade(1996)} Bade N., Komossa S., Dahlem M., 1996, A\&A, 309, L35

\bibitem[\protect\citeauthoryear{Balbus \& Mummery}{2018}]{BalbusMummery(2018)} Balbus S.~A., Mummery A., 2018, MNRAS, 2356

\bibitem[\protect\citeauthoryear{Bardeen \& Petterson}{1975}]{BardeenPetterson(1975)} Bardeen J.~M., Petterson J.~A., 1975, ApJ, 195, L65

\bibitem[\protect\citeauthoryear{Bardeen, Press \& Teukolsky}{1972}]{Bardeen(1972)} Bardeen J.~M., Press W.~H., Teukolsky S.~A., 1972, ApJ, 178, 347

\bibitem[\protect\citeauthoryear{Barker \& Ogilvie}{2014}]{BarkerOgilvie(2014)} Barker A.~J., Ogilvie G.~I., 2014, MNRAS, 445, 2637

\bibitem[\protect\citeauthoryear{Bate et al.}{2000}]{Bate(2000)} Bate M.~R., Bonnell I.~A., Clarke C.~J., Lubow S.~H., Ogilvie G.~I., Pringle J.~E., Tout C.~A., 2000, MNRAS, 317, 773

\bibitem[\protect\citeauthoryear{Bloom et al.}{2011}]{Bloom(2011)} Bloom J.~S., et al., 2011, Sci, 333, 203

\bibitem[\protect\citeauthoryear{Bogdanovi{\'c} et al.}{2004}]{Bogdanovic(2004)} Bogdanovi{\'c} T., Eracleous M., Mahadevan S., Sigurdsson S., Laguna P., 2004, ApJ, 610, 707

\bibitem[\protect\citeauthoryear{Bonnerot et al.}{2016}]{Bonnerot(2016)} Bonnerot C., Rossi E.~M., Lodato G., Price D.~J., 2016, MNRAS, 455, 2253

\bibitem[\protect\citeauthoryear{Brown et al.}{2015}]{Brown(2015)} Brown G.~C., et al., 2015, MNRAS, 452, 4297

\bibitem[\protect\citeauthoryear{Burrows et al.}{2011}]{Burrows(2011)} Burrows D.~N., et al., 2011, Natur, 476, 421

\bibitem[\protect\citeauthoryear{Cao et al.}{2018}]{Cao(2018)} Cao R., Liu F.~K., Zhou Z.~Q., Komossa S., Ho L.~C., 2018, MNRAS, 480, 2929

\bibitem[\protect\citeauthoryear{Cannizzo, Lee \& Goodman}{1990}]{Cannizzo(1990)} Cannizzo J.~K., Lee H.~M., Goodman J., 1990, ApJ, 351, 38

\bibitem[\protect\citeauthoryear{Cenko et al.}{2012a}]{Cenko(2012a)} Cenko S.~B., et al., 2012, ApJ, 753, 77

\bibitem[\protect\citeauthoryear{Cenko et al.}{2012b}]{Cenko(2012b)} Cenko S.~B., et al., 2012, MNRAS, 420, 2684

\bibitem[\protect\citeauthoryear{Chakraborty \& Bhattacharyya}{2017}]{ChakrabortyBhattacharyya(2017)} Chakraborty C., Bhattacharyya S., 2017, MNRAS, 469, 3062

\bibitem[\protect\citeauthoryear{Chan, Krolik \& Piran}{2018}]{Chan(2018)} Chan C.-H., Krolik J.~H., Piran T., 2018, ApJ, 856, 12

\bibitem[\protect\citeauthoryear{Chornock et al.}{2014}]{Chornock(2014)} Chornock R., et al., 2014, ApJ, 780, 44

\bibitem[\protect\citeauthoryear{Curd \& Narayan}{2019}]{CurdNarayan(2018)} Curd B., Narayan R., 2019, MNRAS, 483, 565

\bibitem[\protect\citeauthoryear{Dai et al.}{2018}]{Dai(2018)} Dai L., McKinney J.~C., Roth N., Ramirez-Ruiz E., Miller M.~C., 2018, ApJ, 859, L20

\bibitem[\protect\citeauthoryear{Demianski \& Ivanov}{1997}]{DemianskiIvanov(1997)} Demianski M., Ivanov P.~B., 1997, A\&A, 324, 829

\bibitem[\protect\citeauthoryear{Donato et al.}{2014}]{Donato(2014)} Donato D., et al., 2014, ApJ, 781, 59

\bibitem[\protect\citeauthoryear{Evans \& Kochanek}{1989}]{EvansKochanek(1989)} Evans C.~R., Kochanek C.~S., 1989, ApJ, 346, L13

\bibitem[\protect\citeauthoryear{Foucart \& Lai}{2014}]{FoucartLai(2014)} Foucart F., Lai D., 2014, MNRAS, 445, 1731

\bibitem[\protect\citeauthoryear{Fragile et al.}{2007}]{Fragile(2007)} Fragile P.~C., Blaes O.~M., Anninos P., Salmonson J.~D., 2007, ApJ, 668, 417

\bibitem[\protect\citeauthoryear{Fragile \& Anninos}{2005}]{FragileAnninos(2005)} Fragile P.~C., Anninos P., 2005, ApJ, 623, 347

\bibitem[\protect\citeauthoryear{Fu \& Lai}{2009}]{FuLai(2009)} Fu W., Lai D., 2009, ApJ, 690, 1386

\bibitem[\protect\citeauthoryear{Fuller \& Lai}{2013}]{FullerLai(2013)} Fuller J., Lai D., 2013, MNRAS, 430, 274

\bibitem[\protect\citeauthoryear{Fuller \& Lai}{2012}]{FullerLai(2012)} Fuller J., Lai D., 2012, MNRAS, 421, 426

\bibitem[\protect\citeauthoryear{Franchini, Lodato \& Facchini}{2016}]{Franchini(2016)} Franchini A., Lodato G., Facchini S., 2016, MNRAS, 455, 1946

\bibitem[\protect\citeauthoryear{Gezari et al.}{2006}]{Gezari(2006)} Gezari S., et al., 2006, ApJ, 653, L25

\bibitem[\protect\citeauthoryear{Gezari et al.}{2008}]{Gezari(2008)} Gezari S., et al., 2008, ApJ, 676, 944

\bibitem[\protect\citeauthoryear{Gezari et al.}{2009}]{Gezari(2009)} Gezari S., et al., 2009, ApJ, 698, 1367

\bibitem[\protect\citeauthoryear{Greiner et al.}{2000}]{Greiner(2000)} Greiner J., Schwarz R., Zharikov S., Orio M., 2000, A\&A, 362, L25

\bibitem[\protect\citeauthoryear{Guillochon, Manukian \& Ramirez-Ruiz}{2014}]{Guillochon(2014)} Guillochon J., Manukian H., Ramirez-Ruiz E., 2014, ApJ, 783, 23

\bibitem[\protect\citeauthoryear{Guillochon \& Ramirez-Ruiz}{2013}]{GuillochonRamirez-Ruiz(2013)} Guillochon J., Ramirez-Ruiz E., 2013, ApJ, 767, 25

\bibitem[\protect\citeauthoryear{Guillochon \& Ramirez-Ruiz}{2015}]{GuillochonRamirez-Ruiz(2015)} Guillochon J., Ramirez-Ruiz E., 2015, ApJ, 809, 166

\bibitem[\protect\citeauthoryear{Hawley}{2000}]{Hawley(2000)} Hawley J.~F., 2000, ApJ, 528, 462

\bibitem[\protect\citeauthoryear{Hawley, Guan \& Krolik}{2011}]{Hawley(2011)} Hawley J.~F., Guan X., Krolik J.~H., 2011, ApJ, 738, 84

\bibitem[\protect\citeauthoryear{Hawley \& Krolik}{2018}]{Hawleykrolik(2018)} Hawley J.~F., Krolik J.~H., 2018, ArXiv e-prints, arXiv:1809.01979

\bibitem[\protect\citeauthoryear{Hayasaki, Stone \& Loeb}{2013}]{Hayasaki(2013)} Hayasaki K., Stone N., Loeb A., 2013, MNRAS, 434, 909 

\bibitem[\protect\citeauthoryear{Hayasaki, Stone \& Loeb}{2016}]{Hayasaki(2016)} Hayasaki K., Stone N., Loeb A., 2016, MNRAS, 461, 3760

\bibitem[\protect\citeauthoryear{Ivanov \& Illarionov}{1997}]{IvanovIllarionov(1997)} Ivanov P.~B., Illarionov A.~F., 1997, MNRAS, 285, 394

\bibitem[\protect\citeauthoryear{Ivanov, Zhuravlev \& Papaloizou}{2018}]{Ivanov(2018)} Ivanov P.~B., Zhuravlev V.~V., Papaloizou J.~C.~B., 2018, MNRAS, 2379

\bibitem[\protect\citeauthoryear{Kato}{1990}]{Kato(1990)} Kato S., 1990, PASJ, 42, 99

\bibitem[\protect\citeauthoryear{Khabibullin \& Sazonov}{2014}]{KhabibullinSazonov(2014)} Khabibullin I., Sazonov S., 2014, MNRAS, 444, 1041

\bibitem[\protect\citeauthoryear{King et al.}{2005}]{King(2005)} King A.~R., Lubow S.~H., Ogilvie G.~I., Pringle J.~E., 2005, MNRAS, 363, 49

\bibitem[\protect\citeauthoryear{Komossa \& Bade}{1999}]{KomossaBade(1999)} Komossa S., Bade N., 1999, A\&A, 343, 775

\bibitem[\protect\citeauthoryear{Komossa, et al.}{2008}]{Komossa(2008)} Komossa S., et al., 2008, ApJ, 678, L13

\bibitem[\protect\citeauthoryear{Krolik, et al.}{2016}]{Krolik(2016)} Krolik J., Piran T., Svirski G., Cheng R.~M., 2016, ApJ, 827, 127

\bibitem[\protect\citeauthoryear{Krolik \& Hawley}{2015}]{KrolikHawley(2015)} Krolik J.~H., Hawley J.~F., 2015, ApJ, 806, 141

\bibitem[\protect\citeauthoryear{Kumar \& Pringle}{1985}]{KumarPringle(1985)} Kumar S., Pringle J.~E., 1985, MNRAS, 213, 435

\bibitem[\protect\citeauthoryear{Lei, Zhang \& Gao}{2013}]{Lei(2013)} Lei W.-H., Zhang B., Gao H., 2013, ApJ, 762, 98

\bibitem[\protect\citeauthoryear{Levan et al.}{2011}]{Levan(2011)} Levan A.~J., et al., 2011, Sci, 333, 199

\bibitem[\protect\citeauthoryear{Lightman \& Eardley}{1974}]{LightmanEardly(1974)} Lightman A.~P., Eardley D.~M., 1974, ApJ, 187, L1

\bibitem[\protect\citeauthoryear{Lin et al.}{2015}]{Lin(2015)} Lin D., et al., 2015, ApJ, 811, 43

\bibitem[\protect\citeauthoryear{Liska et al.}{2018a}]{Liska(2018)} Liska M., Hesp C., Tchekhovskoy A., Ingram A., van der Klis M., Markoff S., 2018a, MNRAS, 474, L81

\bibitem[\protect\citeauthoryear{Liska et al.}{2018b}]{Liska(2018b)} Liska M., Tchekhovskoy A., Ingram A., van der Klis M., 2018b, ArXiv e-prints, arXiv:1810.00883

\bibitem[\protect\citeauthoryear{Liu et al.}{2017}]{Liu(2017)} Liu F.~K., Zhou Z.~Q., Cao R., Ho L.~C., Komossa S., 2017, MNRAS, 472, L99

\bibitem[\protect\citeauthoryear{Lodato, King \& Pringle}{2009}]{Lodato(2009)} Lodato G., King A.~R., Pringle J.~E., 2009, MNRAS, 392, 332

\bibitem[\protect\citeauthoryear{Lodato \& Pringle}{2007}]{LodatoPringle(2007)} Lodato G., Pringle J.~E., 2007, MNRAS, 381, 1287

\bibitem[\protect\citeauthoryear{Lodato \& Pringle}{2006}]{LodatoPringle(2006)} Lodato G., Pringle J.~E., 2006, MNRAS, 368, 1196
Calculate the warp and alignment timescale for a SMBH misaligned with the AGN disk.  Neglect pressure torques.

\bibitem[\protect\citeauthoryear{Loeb \& Ulmer}{1997}]{LoebUlmer(1997)} Loeb A., Ulmer A., 1997, ApJ, 489, 573

\bibitem[\protect\citeauthoryear{Lubow, Ogilvie \& Pringle}{2002}]{Lubow(2002)} Lubow S.~H., Ogilvie G.~I., Pringle J.~E., 2002, MNRAS, 337, 706

\bibitem[\protect\citeauthoryear{Lubow \& Ogilvie}{2000}]{LubowOgilvie(2000)} Lubow S.~H., Ogilvie G.~I., 2000, ApJ, 538, 326

\bibitem[\protect\citeauthoryear{Maksym, Ulmer \& Eracleous}{2010}]{Maksym(2010)} Maksym W.~P., Ulmer M.~P., Eracleous M., 2010, ApJ, 722, 1035

\bibitem[\protect\citeauthoryear{Maksym, Lin \& Irwin}{2014}]{Maksym(2014)} Maksym W.~P., Lin D., Irwin J.~A., 2014, ApJ, 792, L29

\bibitem[\protect\citeauthoryear{Martin et al.}{2019}]{Martin(2019)} Martin R.~G., et al., 2019, arXiv e-prints, arXiv:1902.11073

\bibitem[\protect\citeauthoryear{Martin, Pringle \& Tout}{2009}]{Martin(2009)} Martin R.~G., Pringle J.~E., Tout C.~A., 2009, MNRAS, 400, 383

\bibitem[\protect\citeauthoryear{Metzger \& Stone}{2016}]{MetzgerStone(2016)} Metzger B.~D., Stone N.~C., 2016, MNRAS, 461, 948

\bibitem[\protect\citeauthoryear{Montesinos Armijo \& de Freitas Pacheco}{2011}]{Montesinos(2011)} Montesinos Armijo M., de Freitas Pacheco J.~A., 2011, ApJ, 736, 126

\bibitem[\protect\citeauthoryear{Morales Teixeira et al.}{2014}]{Morales(2014)} Morales Teixeira D., Fragile P.~C., Zhuravlev V.~V., Ivanov P.~B., 2014, ApJ, 796, 103

\bibitem[\protect\citeauthoryear{Nealon et al.}{2016}]{Nealon(2016)} Nealon R., Nixon C., Price D.~J., King A., 2016, MNRAS, 455, L62

\bibitem[\protect\citeauthoryear{Nelson \& Papaloizou}{2000}]{NelsonPapaloizou(2000)} Nelson R.~P., Papaloizou J.~C.~B., 2000, MNRAS, 315, 570

\bibitem[\protect\citeauthoryear{Ogilvie \& Lynch}{2019}]{OgilvieLynch(2019)} Ogilvie G.~I., Lynch E.~M., 2019, MNRAS, 483, 4453

\bibitem[\protect\citeauthoryear{Ogilvie}{2001}]{Ogilvie(2001)} Ogilvie G.~I., 2001, MNRAS, 325, 231

\bibitem[\protect\citeauthoryear{Ogilvie}{1999}]{Ogilvie(1999)} Ogilvie G.~I., 1999, MNRAS, 304, 557

\bibitem[\protect\citeauthoryear{Papaloizou \& Lin}{1995}]{PapaloizouLin(1995)} Papaloizou J.~C.~B., Lin D.~N.~C., 1995, ApJ, 438, 841

\bibitem[\protect\citeauthoryear{Papaloizou \& Pringle}{1983}]{PapaloizouPringle(1983)} Papaloizou J.~C.~B., Pringle J.~E., 1983, MNRAS, 202, 1181

\bibitem[\protect\citeauthoryear{Piran et al.}{2015}]{Piran(2015)} Piran T., Svirski G., Krolik J., Cheng R.~M., Shiokawa H., 2015, ApJ, 806, 164

\bibitem[\protect\citeauthoryear{Press et al.}{2002}]{Press(2002)} Press W.~H., Teukolsky S.~A., Vetterling W.~T., Flannery B.~P., 2002, Numerical recipes in C++ : the art of scientific computing

\bibitem[\protect\citeauthoryear{Rees}{1988}]{Rees(1988)} Rees M.~J., 1988, Natur, 333, 523

\bibitem[\protect\citeauthoryear{Saxton et al.}{2012}]{Saxton(2012b)} Saxton C.~J., Soria R., Wu K., Kuin N.~P.~M., 2012, MNRAS, 422, 1625

\bibitem[\protect\citeauthoryear{Scheuer \& Feiler}{1996}]{ScheuerFeiler(1996)} Scheuer P.~A.~G., Feiler R., 1996, MNRAS, 282, 291

\bibitem[\protect\citeauthoryear{Shen \& Matzner}{2014}]{ShenMatzner(2014)} Shen R.-F., Matzner C.~D., 2014, ApJ, 784, 87

\bibitem[\protect\citeauthoryear{Shiokawa et al.}{2015}]{Shiokawa(2015)} Shiokawa H., Krolik J.~H., Cheng R.~M., Piran T., Noble S.~C., 2015, ApJ, 804, 85

\bibitem[\protect\citeauthoryear{Sorathia, Krolik \& Hawley}{2013}]{Sorathia(2013)} Sorathia K.~A., Krolik J.~H., Hawley J.~F., 2013, ApJ, 777, 21

\bibitem[\protect\citeauthoryear{Stone \& Loeb}{2012}]{StoneLoeb(2012)} Stone N., Loeb A., 2012, PhRvL, 108, 61302

\bibitem[\protect\citeauthoryear{Strubbe \& Quataert}{2009}]{StrubbeQuataert(2009)} Strubbe L.~E., Quataert E., 2009, MNRAS, 400, 2070

\bibitem[\protect\citeauthoryear{Tchekhovskoy et al.}{2014}]{Tchekhovskoy(2014)} Tchekhovskoy A., Metzger B.~D., Giannios D., Kelley L.~Z., 2014, MNRAS, 437, 2744

\bibitem[\protect\citeauthoryear{Tremaine \& Davis}{2014}]{TremaineDavis(2014)} Tremaine S., Davis S.~W., 2014, MNRAS, 441, 1408

\bibitem[\protect\citeauthoryear{van Velzen et al.}{2018a}]{vanVelzen(2018)} van Velzen S., Stone N.~C., Metzger B.~D., Gezari S., Brown T.~M., Fruchter A.~S., 2018, ArXiv e-prints, arXiv:1809.00003

\bibitem[\protect\citeauthoryear{van Velzen et al.}{2011}]{vanVelzen(2011)} van Velzen S., et al., 2011, ApJ, 741, 73

\bibitem[\protect\citeauthoryear{Wang et al.}{2011}]{Wang(2011)} Wang T.-G., Zhou H.-Y., Wang L.-F., Lu H.-L., Xu D., 2011, ApJ, 740, 85

\bibitem[\protect\citeauthoryear{Wang et al.}{2012}]{Wang(2012)} Wang T.-G., Zhou H.-Y., Komossa S., Wang H.-Y., Yuan W., Yang C., 2012, ApJ, 749, 115

\bibitem[\protect\citeauthoryear{Wienkers \& Ogilvie}{2018}]{WienkersOgilvie(2018)} Wienkers A.~F., Ogilvie G.~I., 2018, MNRAS, 477, 4838

\bibitem[\protect\citeauthoryear{Xiang-Gruess, Ivanov \& Papaloizou}{2016}]{XiangGruess(2016)} Xiang-Gruess M., Ivanov P.~B., Papaloizou J.~C.~B., 2016, MNRAS, 463, 2242

\bibitem[\protect\citeauthoryear{Zauderer et al.}{2011}]{Zauderer(2011)} Zauderer B.~A., et al., 2011, Natur, 476, 425

\bibitem[\protect\citeauthoryear{Zhuravlev et al.}{2014}]{Zhuravlev(2014)} Zhuravlev V.~V., Ivanov P.~B., Fragile P.~C., Morales Teixeira D., 2014, ApJ, 796, 104

\bibitem[\protect\citeauthoryear{Zhuravlev \& Ivanov}{2011}]{ZhuravlevIvanov(2011)} Zhuravlev V.~V., Ivanov P.~B., 2011, MNRAS, 415, 2122

\end{thebibliography}
\end{document}